\documentclass[preprint]{aastex}

\newcommand{\ug}{\hbox{$u\!-\!g$}}		
\newcommand{\gr}{\hbox{$g\!-\!r$}}
\newcommand{\ri}{\hbox{$r\!-\!i$}}
\newcommand{\bvt}{\hbox{$B_T\!-\!V_T$}}
\newcommand{\ugriz}{\hbox{$u\,g\,r\,i\,z$}}
\newcommand{\ugiz}{\hbox{$u\,g\,i\,z$}}
\newcommand{\iuzg}{\hbox{$i\,u\,z\,g$}}
\newcommand{\ssc}{{\tt SSC}}
\newcommand{\da}{{\tt DA}}
\newcommand{\psp}{{\tt PSP}}
\newcommand{\astrom}{{\tt Astrom}}
\newcommand{\frames}{{\tt Frames}}

\shorttitle{SDSS Astrometric Calibration}
\shortauthors{Pier et al.}

\begin{document}

\title{Astrometric Calibration of the Sloan Digital Sky Survey}

\author{Jeffrey R. Pier and Jeffrey A. Munn}
\affil{US Naval Observatory, Flagstaff Station, P.O. Box 1149, Flagstaff,
AZ 86002-1149; jrp@nofs.navy.mil, jam@nofs.navy.mil}

\author{Robert B. Hindsley}
\affil{Remote Sensing Division, Code 7215, Naval Research Laboratory,
4555 Overlook Avenue, SW, Washington, DC 20375; hindsley@nrl.navy.mil}

\author{G. S. Hennessy}
\affil{US Naval Observatory, 3450 Massachusetts Avenue, NW, Washington, DC
20392-5420; gsh@usno.navy.mil}

\author{Stephen M. Kent}
\affil{Fermi National Accelerator Laboratory, P.O. Box 500, Batavia, IL
60510; skent@fnal.gov}

\and

\author{Robert H. Lupton and \v{Z}eljko Ivezi\'{c}}
\affil{Princeton University Observatory, Princeton, NJ 08544;
rhl@astro.princeton.edu, ivezic@astro.princeton.edu}

\begin{abstract}

The astrometric calibration of the Sloan Digital Sky Survey is
described.  For point sources brighter than $r \sim 20$ the
astrometric accuracy is 45 milliarcseconds (mas) rms per coordinate
when reduced against the USNO CCD Astrograph Catalog, and 75 mas rms when
reduced against Tycho-2, with an additional 20 -- 30 mas systematic error
in both cases.
The rms errors are dominated by anomalous refraction and random errors in
the primary reference catalogs.
The relative astrometric accuracy between the $r$ filter and each
of the other filters (\ugiz) is 25 -- 35 mas rms.
At the survey limit ($r \sim 22$), the astrometric accuracy is limited by
photon statistics to approximately 100 mas rms for typical seeing.
Anomalous refraction is shown to contain components correlated over two or
more degrees on the sky.

\end{abstract}

\keywords{astrometry --- methods: data analysis --- surveys}

\section{Introduction}

The Sloan Digital Sky Survey (SDSS) is obtaining images of one-quarter of the
sky in five broad optical bands
\citep[$u$, $g$, $r$, $i$, and $z$;][]{filters,smith,hogg},
with 95\% completeness limits for point sources of
22.0, 22.2, 22.2, 21.3, and 20.5, respectively.  The survey will produce a
database of roughly $10^8$ galaxies, $10^8$ stars, and $10^6$ QSOs
with accurate photometry, astrometry, and
object classification parameters \citep{sdss}.  Spectroscopy, covering
the wavelength range 3800\,\AA\ to 9200\,\AA\ at $R \approx 1800$, will be
obtained for roughly 1,000,000 galaxies, 100,000 quasars, and another 50,000
stars and serendipitous objects.  (Currently, funding is available to complete
approximately two-thirds of the survey area.)
At the time of this writing,
approximately 3600 square degrees of sky have been imaged and spectra have
been obtained for 400,000 objects.

This paper describes the astrometric calibration of the SDSS.
Accurate placement
of fibers for the survey spectroscopy requires an accuracy for
astrometry of 180 milliarcseconds (mas) rms per coordinate (all errors and
accuracies quoted in this paper are rms per coordinate).
In fact, the final accuracy is
considerably better.  Depending on the reference catalog available in a given
area of sky, for point sources brighter than
$r \sim 20$ the astrometry is accurate to 45 mas rms when reduced against
the United States Naval Observatory CCD Astrograph Catalog
\citep[UCAC,][]{ucac},
and 75 mas rms when reduced against Tycho-2 \citep{tycho2}, with additional
systematic errors of 20 -- 30 mas in both cases.
The accuracy of the relative astrometry between the $r$
detection and the $u$, $g$, $i$, and $z$ detections is typically 25 -- 35 mas.

Section~\ref{sec-camera} describes the layout of the imaging camera, which
motivates much of the astrometric calibration process.
Section~\ref{sec-overview} gives an overview of the imaging data processing
pipelines, providing a context for the detailed discussion of the Astrometric
Pipeline.  Section~\ref{sec-centroids} discusses centroiding
algorithms.  The Astrometric Pipeline and calibration procedures are described
in Section~\ref{sec-calibrations}, and the accuracy of the resultant astrometry
is assessed in Section~\ref{sec-results}.  Differences between the calibrations
described here and those employed in previously released SDSS data are
described in Section~\ref{sec-edr}.  A few concluding remarks are given in
Section~\ref{sec-conclusion}.

\section{Imaging Camera}
\label{sec-camera}

\begin{figure}
\plotone{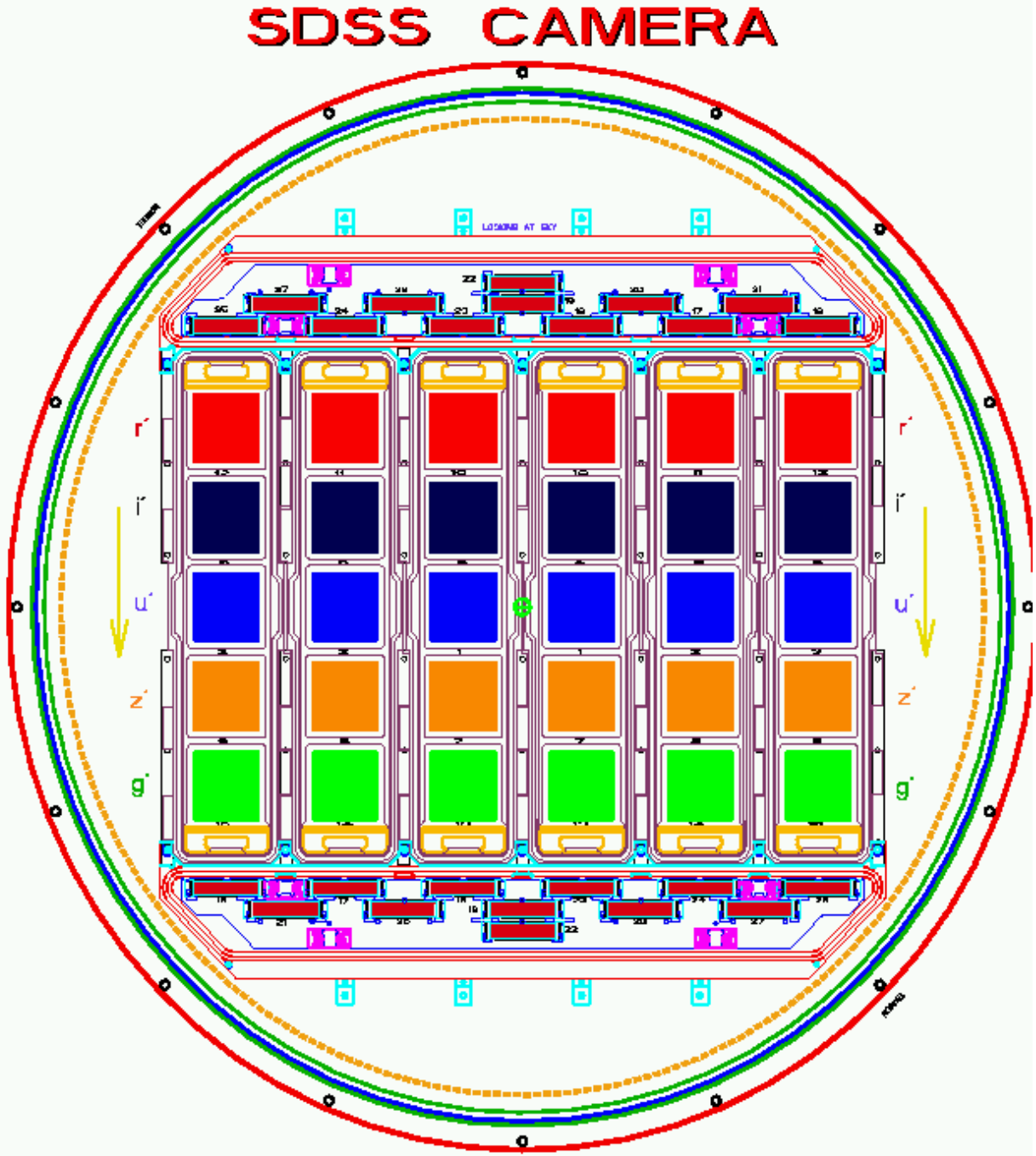}
\caption{Layout of the Imaging Camera.  The large colored squares represent
the photometric CCDs, five in each of six dewars which run vertically on the
page.  The smaller rectangles above and below show the 22 astrometric and two
focus CCDs in two dewars running horizontally.  The arrows indicate the
direction of motion of stars across the camera during a scan.}
\label{fig-camera}
\end{figure}

The imaging camera \citep[Figure~\ref{fig-camera}]{camera} consists of 54
CCDs in eight dewars and spans $2.3\arcdeg$ on the sky.  Thirty of these CCDs
are the main imaging/photometric devices, each a SITe (Scientific
Imaging Technologies, formerly Tektronix) device with 2048$^2$ -- 24$\mu$m
square pixels.
They are arranged in six column dewars of five CCDs each, one CCD for
each filter bandpass (\ugriz) in each column.  The camera is
operated in TDI (Time Delay and Integrate), or scanning, mode for which
the telescope
is driven at a rate synchronous with the charge transfer rate of
the CCDs.  Objects on the sky drift down columns of the CCDs so
that five-color photometry is obtained over an interval of 5 minutes with no
dead-time for shuttering and read-out.
The effective integration time is 53.9 seconds at the chosen (sidereal)
scanning rate (the pixel scale is 0.396 arcseconds per pixel).

The other 24 CCDs in two additional dewars are also SITe chips of width
2048 -- 24$\mu$m pixels,
but they have only 400 rows in the scanning direction.  These dewars are
oriented perpendicular to the photometric dewars, and one lies across the
top of the columns, the other across the bottom. Each dewar contains 12
CCDs. Two of these CCDs (one in each dewar)
are used to dynamically determine changes in focus.  The other 22 CCDs are
used to tie observations to an astrometric reference frame.  They
saturate at brighter magnitudes than the photometric CCDs  because they
(1) have a shorter integration time (10.5 seconds), since they have 400
versus 2048 rows in the scanning direction; (2) are behind a neutral density
filter of 3.0 stellar magnitudes; and (3) have somewhat poorer quantum
efficiency.
The astrometric CCD filter bandpass is similar to the photometric $r$
bandpass, though it extends $\sim 400$\,\AA\ redwards of the $r$ red cutoff.
The astrometric CCDs saturate at $r \simeq 8$,
and thus can utilize all but the brightest 10\% of the
astrometric standard stars in Tycho-2.  The photometric $r$
CCDs saturate at $r \simeq 14$, while good centroids may be measured on
the astrometric CCDs for stars as faint as $r \simeq 17$.
Hence there is an overlap of about 3 magnitudes between faint stars
with good centroids on the astrometric chips and bright unsaturated stars
on the $r$ chips, enabling the transfer of catalog stars detected on the
astrometric CCDs to the $r$ CCDs, and thus the astrometric calibration of
stars detected on the $r$ CCDs
against bright reference catalogs.  Similarly, there is an overlap of at
least 5 magnitudes
between stars with good centroids detected on the $r$ CCDs and the
other photometric CCDs, allowing the $r$ astrometric calibrations to be
transferred to the other photometric CCDs.

Twelve of the astrometric CCDs are aligned with the six columns of
photometric CCDs, with an astrometric chip leading and trailing each
column.  The other 10 astrometric chips bridge the gaps between
the columns.  Hence the astrometric chips span the full 2.3$\arcdeg$ width
of the camera.  As outlined below, the bridge astrometric CCDs are not used in
the current version of the astrometric pipeline.  They were originally intended
to account for differential shifts between the photometric dewars, but
the camera has proven to be extremely stable and the technique was not needed.

\section{Data Processing Overview}
\label{sec-overview}

The large volume of data generated by the SDSS requires a set of automated
data processing pipelines to acquire and archive the raw data,
detect and measure object parameters, astrometrically and
photometrically calibrate the objects, select spectroscopic targets,
design the spectroscopic plates, extract spectra, and measure various
spectral parameters.  The end-to-end processing of SDSS data is discussed
in detail in \citet{edr}.  Those parts of data
processing which are relevant
to the astrometric calibrations are briefly summarized here to give context
to the detailed discussion of the astrometric calibrations.

Each imaging drift scan is processed serially through the following pipelines:

\begin{enumerate}

\item The real-time Data Acquisition System \citep[\da,][]{da} acquires
the raw data as they are read off the CCDs, buffers them to disk, archives
them to tape, and performs
real-time quality analysis (it is the only real-time component of data
processing).  The data stream from each CCD is broken up into frames
(2048 columns by 1361 rows) to facilitate data processing.
\da\ finds bright stars on the frames from the astrometric CCDs, cuts out
a 29 by 29 pixel (11.5 by 11.5 arcseconds) subraster ({\em postage stamp})
centered on each star,
and measures parameters, including a centroid and flux, for each star
(the astrometric frames are not saved).

\item The Serial Stamp Collection Pipeline \citep[\ssc,][]{photo} cuts out
postage
stamps and measures centroids for bright stars on the photometric CCD frames.
It measures stars at the expected positions based on the \da\ detections on the
astrometric CCDs in the same camera column and includes a star finder to
detect additional stars not
detected on the astrometric CCDs.

\item The Postage Stamp Pipeline \citep[\psp,][]{photo} uses postage stamps for
bright stars cut out by \ssc\ to measure the point spread function (PSF) as a
function of position across each frame (for the photometric CCDs only).
The star centroids for all postage stamps are remeasured using an algorithm
that corrects
for asymmetries in the PSF, as described in Section~\ref{sec-centroids}.

\item The Astrometric Pipeline (\astrom) matches up the stars detected by
\da\ for the astrometric CCDs, and/or those detected by \ssc\ for the
photometric CCDs with corrected centroids produced by \psp, to a primary
astrometric reference catalog.  Using these matches, it derives a
separate set of transformation equations for each frame, converting frame
row and column to J2000 catalog mean place celestial coordinates.  This
procedure provides the astrometric calibration for the SDSS, and is the topic
of this paper.
The pipeline itself is described in detail in Section~\ref{sec-calibrations}.

\item The Imaging Pipeline \citep[\frames,][]{photo} corrects the raw frames
for instrumental effects,
detects objects on the corrected frames, matches up detections of the same
object in different filters, deblends overlapping objects, and measures
parameters for the objects.  It uses the astrometric calibrations
produced by \astrom\ to precisely match up the separate detections in the
different filters for each object.  This requires that \astrom\ run before
\frames, and explains why the astrometric
calibrations are based on centroids measured by \psp\ and \da, rather than on
\frames's centroids.  \frames\ and \psp\ use the same centroiding algorithm.
The final SDSS object positions are derived by applying the astrometric
calibrations produced by \astrom\ to the centroids measured by \frames.

\end{enumerate}

In summary, the astrometric calibrations are produced by \astrom, processing a
single drift scan at a time, using centroids measured by \da\ and/or \psp\ and
producing a set of transformation equations for each frame which are applied
to centroids measured by \frames.

\section{Centroids}
\label{sec-centroids}

An object's centroid is defined as the first moment of its light distribution.
Since this is a noisy estimate for most objects in the survey, the following
technique is used to better estimate the centroid.  First, an object's
image is smoothed using a two-dimensional Gaussian with an adaptive smoothing
length scaled to the PSF in that frame.
Quartic interpolation is used to find the maximum in a $3 \times 3$ pixel
subraster centered on the peak pixel in the smoothed image.  This gives a
biased estimate of the centroid in the presence of an asymmetric PSF.
The Postage Stamp Pipeline (\psp) uses bright stars to determine the shape of
the PSF as a smoothly varying function of CCD column and row
for each frame.  Thus, the PSF at the position of each object
is determined to high accuracy.  The centroid of the PSF at the
position of the object is measured using both the first moment
and quartic interpolation after smoothing by the PSF.  The difference
in centroids measured for the PSF using the two algorithms is a measure
of the bias introduced by the use of the quartic interpolation algorithm.
This bias is typically a few tens of mas, but can be as large as 100 mas.  The
difference is added to the quartic interpolation centroid measured for the
object, yielding a high signal-to-noise ratio estimate of the first moment
centroid of the object.  A more detailed discussion of the centroiding
algorithm will appear in the forthcoming paper on SDSS photometry
(R. Lupton et al., in preparation).

\psp\ and \frames\ both use the bias-corrected centroiding algorithm described
above.  This is necessary because the
astrometric calibrations are produced from centroids measured by \psp\
but are applied to centroids measured by \frames.  Since neither \da\ nor \ssc\
has an estimate of the PSF, they measure uncorrected quartic interpolation
centroids, using a constant smoothing length of $\sigma = 1.2$ pixels
($\sim 0.48$ arcseconds).
When reducing against Tycho-2, \astrom\ uses \da's uncorrected centroids for
objects detected on the astrometric CCDs; however the biases thus introduced
are largely removed in the calibrations and are small compared to more dominant
atmospheric effects.

In addition to its row and column centroid, each detection of a star is
characterized by the terrestrial time (TT) when the star
was at mid-exposure, its flux on that frame, and
an instrumental color based on detections on two CCDs in the same camera column
during the same drift scan (\ri\ for the $r$, $i$, $z$, and astrometric
CCDs, \ug\ for the $u$ CCDs, and \gr\ for the $g$ CCDs).
\astrom\ converts instrumental fluxes and colors
to approximate calibrated magnitudes and colors using a separate zero
point for each CCD (derived from a recent photometric calibration), the
airmass at the middle of the frame, and average values
for the extinction (0.520 mags per unit airmass in $u$, 0.200 in $g$, 0.120
in $r$, 0.080 in $i$,
and 0.065 in $z$).  These calibrated magnitudes are used by \astrom\ for quality
analysis and to reject faint stars.  The calibrated colors are used for
differential chromatic refraction (DCR) calculations.  The expected errors in
the approximate magnitudes
and colors do not contribute significantly to the overall astrometric
error budget.
Only unsaturated stars brighter than $20^{th}$ magnitude are used to derive
the astrometric calibrations.


\section{Calibrations}
\label{sec-calibrations}

The $r$ CCDs serve as the astrometric reference CCDs for the SDSS.  That is,
the positions for SDSS objects are based on the $r$ centroids and calibrations.
One of two reduction strategies is employed, depending on the coverage of
astrometric catalogs:

\begin{enumerate}
\item Whenever possible, stars detected on the $r$ photometric CCDs are matched
directly with stars in UCAC.  UCAC extends down to $R = 16$, giving
approximately 2 -- 3 magnitudes of overlap with unsaturated stars on the
photometric $r$ CCDs.  The astrometric CCDs are not used.
\item If UCAC coverage is not available to reduce a given imaging scan,
detections of Tycho-2 stars on the astrometric CCDs are mapped onto the
$r$ CCDs (all Tycho-2 stars saturate on the $r$ CCDs) using bright stars that
have sufficient signal-to-noise ratio on the astrometric CCDs and are
unsaturated on the $r$ CCDs. (This mode was originally expected to be the norm;
at the time the camera was designed, UCAC was several years away from beginning
observations.)
\end{enumerate}

The $r$ CCDs are therefore calibrated directly against the primary astrometric
reference catalog.
\frames\ uses the astrometric calibrations to match up
detections of the same object observed in the other four filters.
The accuracy of the
relative astrometry between filters can thus significantly impact \frames,
in particular the deblending of overlapping objects, photometry based on the
same
aperture in different filters, and detection of moving objects.  To minimize
the errors in the relative astrometry between filters, the $u$, $g$, $i$, and
$z$ CCDs are calibrated against the $r$ CCDs.

Each drift scan is processed separately.  All six camera columns are processed
in a single reduction.  In brief, stars detected on the $r$ CCDs if
calibrating against UCAC, or stars detected on the astrometric CCDs transformed
to $r$ coordinates if calibrating against Tycho-2, are matched to catalog
stars.
Transformations from $r$ pixel coordinates to catalog mean place (CMP)
celestial coordinates are derived using
a running-means least-squares fit to a focal plane model,
using all six $r$ CCDs together to solve for both the telescope tracking and
the $r$ CCDs' focal plane offsets, rotations, and scales,
combined with smoothing spline fits to the intermediate residuals.
These transformations, comprising the calibrations for the $r$ CCDs, are then
applied to the stars detected on the $r$ CCDs, converting them to CMP
coordinates and creating a catalog of secondary astrometric standards.  Stars
detected on the $u$, $g$, $i$, and $z$ CCDs are then matched to this
secondary
catalog, and a similar fitting procedure (each CCD is fitted separately)
is used to derive transformations
from the pixel coordinates for the other photometric CCDs to CMP celestial
coordinates, comprising the calibrations for the $u$, $g$, $i$, and $z$ CCDs.
The remainder of this section describes the calibration procedure in detail.

\subsection{Great Circle Coordinates}

In order to minimize curvature and time differences as objects drift
across the focal plane, all SDSS drift scans are conducted along CMP J2000
great circles.  Each such great circle (referred to as a {\em stripe}) is
defined by the inclination and the right ascension of the ascending node
where the great circle crosses the J2000 celestial equator
(\citeauthor{edr} [2002] provides more information on the various
coordinate systems employed in the survey).
The {\em boresight} is defined as the position in the focal
plane that nominally tracks the great circle.  A minimum of two scans is
required along each great circle, one with the boresight offset 22.74~mm
below (perpendicular to the scan direction) the camera center (referred to
as the north {\em strip} of the stripe), and one with the boresight offset
22.74~mm above the camera center (referred to as the south strip).
The two strips are interlaced to fill in the gaps between the
camera columns, such that two strips together cover a swath about
$2.53\arcdeg$ wide centered along the great circle.  The offsets between
strips are chosen so that the individual {\em scan lines} from each of the
six CCD columns overlap slightly ($\sim 0.5 '$) with adjacent scan lines
when interlaced.  Adjacent stripes, separated by $2.5\arcdeg$, also overlap
at their edges.
The boresight tracking rate is adjusted by the telescope control computer (TCC)
to be constant in observed place.

Reductions are simplified by working in a coordinate system in which the
tracked great circle is the equator of the coordinate system.  In this
coordinate system, referred to throughout as the {\em great circle coordinate
system}, the latitude of an observed star never exceeds about $1.3 \arcdeg$;
thus the small angle approximation may be used and lines of
constant longitude are to a high approximation perpendicular to lines of
constant latitude.  Longitude and latitude in great circle coordinates are
referred to with the symbols $\mu$ and $\nu$, respectively.  $\nu = 0$ along
the great circle, $\mu$ increases in the scan direction, and the origin of
$\mu$ is chosen so that $\mu = \alpha_{2000}$ at the ascending node (where
the great circle crosses the J2000 celestial equator).
The conversion from great circle coordinates to J2000 celestial coordinates
is then
\begin{eqnarray}
tan(\alpha_{2000}-\mu_0) & = & [\sin(\mu - \mu_0) \cos \nu \cos i -
                    \sin \nu \sin i] / [\cos(\mu - \mu_0) \cos \nu] \\
sin(\delta_{2000}) & = & \sin (\mu - \mu_0) \cos \nu \sin i +
                    \sin \nu \cos i
\end{eqnarray}
where $i$ and $\mu_0$ are the inclination and J2000 right ascension of the
great circle ascending node, respectively
($\mu_0 = 95\arcdeg$ for all survey stripes
\footnote{A few early commissioning scans, and some
ongoing calibration scans, track great circles which are not survey stripes}).

\subsection{Pseudo-Catalog Place}

As stated above, the telescope boresight tracks (or should track) a J2000
great circle in CMP.  The distance of a star's image from the
boresight on the
telescope focal plane is proportional to the difference in observed place
between the star and the boresight pointing.
The reductions are therefore performed in
{\em pseudo-catalog place} (PCP), a nonstandard place defined as
\begin{eqnarray}
\mu_{\ast,PCP} & = & \mu_{B,CMP} + (\mu_{\ast,OP} - \mu_{B,OP}) \\
\nu_{\ast,PCP} & = & \nu_{B,CMP} + (\nu_{\ast,OP} - \nu_{B,OP})
\end{eqnarray}
where the subscripts $\ast$ and {\em B} indicate the star and boresight,
respectively, and the subscripts {\em PCP}, {\em CMP}, and {\em OP} signify
pseudo-catalog place, catalog mean place, and observed place, respectively
(for definitions of various astrometric places,
see \citeauthor{almanac} [1992]).
Local place could just as well have been used; however using pseudo-catalog
place facilitates the generation of quality analysis information regarding
the telescope tracking and camera rotation over the course of a scan.
That is, the pipeline explicitly solves for parameters describing the actual
telescope scanning.

\subsection{Calibration Equations}
\label{sec-equations}

While the reductions are performed in PCP, ultimately a
transformation from pixel coordinates to CMP is required.
Throughout all SDSS data processing, the data from each CCD in a drift scan
are broken up into contiguous frames of 2048 columns (columns are
parallel to the scan direction) by 1361 rows
(perpendicular to the scan direction).
Astrometric calibrations are generated as a separate set of equations for
each frame converting frame row ($x$), frame column ($y$), and
star color to CMP great circle coordinates
($\mu_{CMP}$, $\nu_{CMP}$).
\begin{eqnarray}
\nonumber
{\rm for}\ \ri & < & (\ri)_0: \\
& & x' = x + g_0 + g_1 y + g_2 y^2 + g_3 y^3
            + p_x color \\
& & y' = y + h_0 + h_1 y + h_2 y^2 + h_3 y^3
            + p_y color \\
\nonumber
{\rm for}\ \ri & \geq & (\ri)_0: \\
& & x' = x + g_0 + g_1 y + g_2 y^2 + g_3 y^3 + q_x \\
& & y' = y + h_0 + h_1 y + h_2 y^2 + h_3 y^3 + q_y \\
\mu_{CMP} & = & a + b x' + c y' \\
\nu_{CMP} & = & d + e x' + f y'
\end{eqnarray}
The transformation from ($x$, $y$) to ($x'$, $y'$) corrects for optical
distortions (which, in TDI mode, are a function of column
only) and DCR.  For $u$ and
$g$ frames, DCR is modeled as a linear function of color (\ug\ for $u$
frames, \gr\ for $g$ frames) for stars bluer than $(\ri)_0 = 1.5$,
and a constant for redder stars. For $r$, $i$, and $z$ frames, DCR is
modeled as a linear function of color (\ri) for all stars
($[\ri]_0 = \infty$).
The corrected frame coordinates ($x'$, $y'$) are then transformed to CMP
great circle coordinates ($\mu_{CMP}$, $\nu_{CMP}$) using an affine
transformation.

\begin{figure}
\plotone{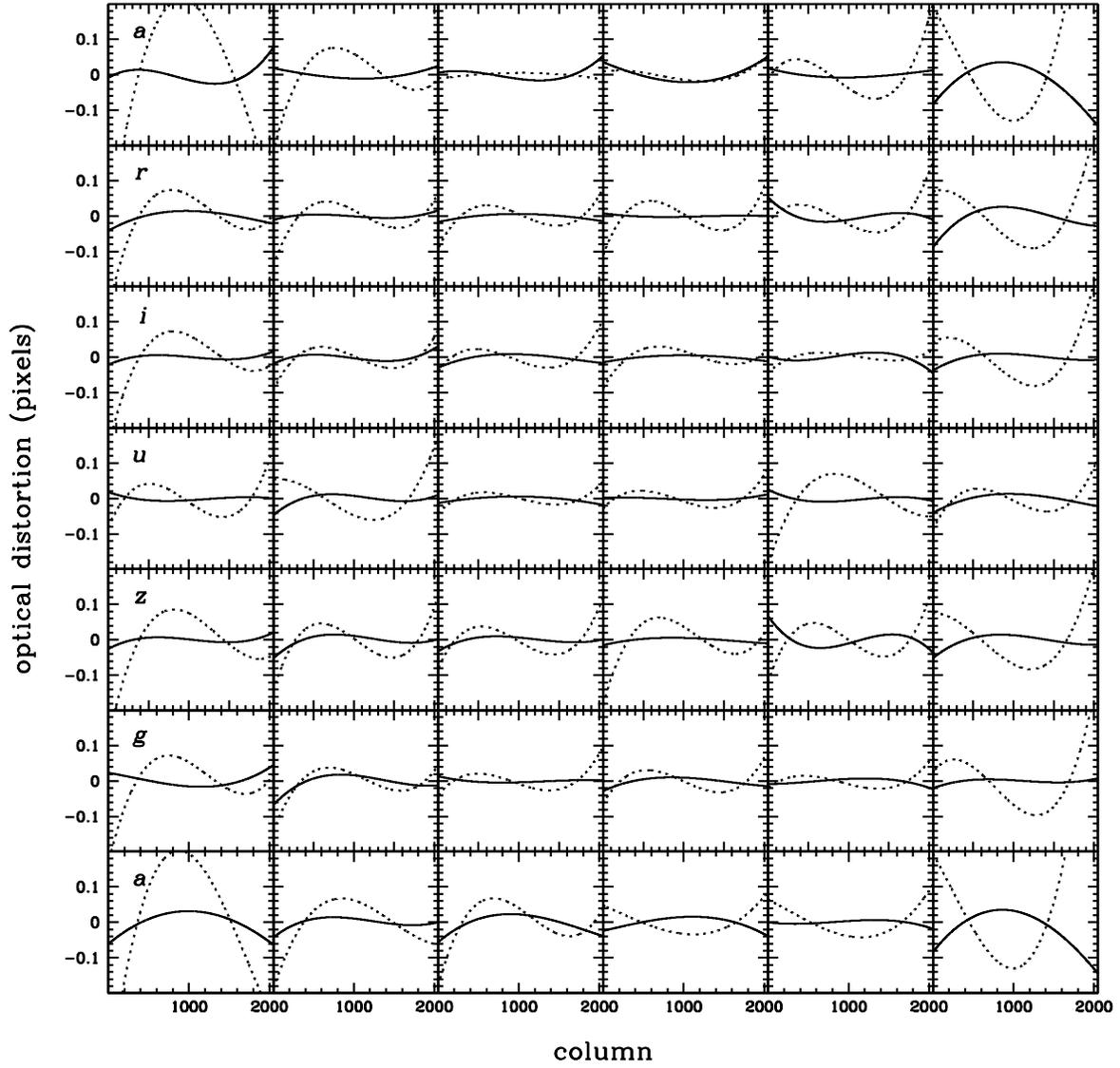}
\caption{Cubic polynomial fits to the observed optical distortions for TDI
scanning, in pixels, as a function of CCD column.
Plots are displayed for each CCD (labelled by filter) in camera
columns 1 -- 6 (left to right) of the imaging camera.
The distortions along the rows are plotted as solid lines, and along the
columns as dotted lines.}
\label{fig-distortions}
\end{figure}

Figure~\ref{fig-distortions} displays cubic polynomial fits to the observed
optical distortions as a function of CCD column for each CCD.
Distortions are typically less than 0.2 pixels,
and are stable from month to month.

The use of a set of transformation equations per frame, rather than directly
calibrating each object, facilitates compartmentalization in data processing
as well as later recalibrations.  However, it can introduce systematic errors
if the equations do not adequately map the true behavior.  After fitting the
optical distortion terms, there remain systematics as a function of column,
though typically of order 5 mas or less.  It is more difficult to estimate
the errors introduced by the approximation that the astrometry changes
linearly with time (that is, with row) over a single frame on the photometric
CCDs; however
they seem to be less than order 10 mas.  (This is not a good approximation for
the astrometric CCDs, for which the effects of seeing are more severe due to
the shorter integration times.)

The equations for fitting DCR were determined
by convolving spectrophotometry of stars from the \citet{gunn-stryker}
spectral atlas
with the SDSS filter curves and a model atmosphere. For the $r$, $i$,
and $z$ filters the linear fits of DCR versus color at a given airmass are
excellent over all spectral types, with rms errors of less than
5 mas at a zenith distance of 60 degrees (the extreme zenith distance at which
observations are obtained; as zenith distance is varied, the DCR errors
scale with refraction).  In $u$ and $g$ the linear fits are poorer but
adequate if confined to stars bluer than $r - i \simeq 1.5$, with rms errors
of 38 and 13 mas, respectively, at a zenith distance of 60 degrees.  For redder
stars the DCR terms in $u$ and $g$ are poorly behaved, and thus a constant
value is employed.  The accuracy of the astrometry, including the
systematics introduced by the form of the transformation equations, is
discussed in Section~\ref{sec-results}.

\subsection{r Calibrations}

The technique used to calibrate the $r$ CCDs depends on the primary standard
star catalog used.  The preferred catalog is
UCAC\footnote{\url{http://ad.usno.navy.mil/ucac/}}, an (eventually) all-sky
astrometric catalog with a precision of 70 mas at its catalog limit
of $R \simeq 16$, and systematic errors of less than 30 mas.
Currently the entire Southern Hemisphere has been observed,
with partial coverage in the north (the entire sky should be finished by
the end of 2003); one-third of the SDSS area is covered at present.  There
are 2 -- 3 magnitudes of overlap between the faint end of UCAC and bright
unsaturated stars on the $r$ CCDs (depending on the CCD and on seeing),
allowing the direct calibration of the $r$ CCDs against UCAC,
bypassing the astrometric CCD to photometric CCD bootstrap.

If a scan is not covered by the current UCAC catalog, then it is reduced
against Tycho-2, an all-sky
astrometric catalog with a median precision of 70 mas at its catalog limit
of $V_T \simeq 11.5$, and systematic errors of less than 1 mas.
All Tycho-2 stars are saturated on the $r$ CCDs;
however there are about 3.5 magnitudes of overlap between bright unsaturated
stars on the astrometric CCDs and the faint end of Tycho-2
($8 < r < 11.5$), and about 3 magnitudes of overlap between bright unsaturated
stars on the $r$ CCDs and faint stars on the astrometric CCDs ($14 < r < 17$).
For reductions against Tycho-2, the star centroids on the astrometric CCDs
are transformed to the pixel coordinates of the $r$ CCD in the same column
using the unsaturated stars detected on both the astrometric and $r$ CCDs.
Each pair of astrometric-$r$ CCDs is processed separately.
Stars on each astrometric CCD are matched to stars on its associated $r$
CCD.
There are typically 10 -- 20 matched pairs per frame.
For each frame, the nearest (in $\mu$) 30 matched pairs, including all
matched pairs on that frame, are used to derive an affine transformation
from the astrometric
CCD pixel coordinate system to the $r$ CCD pixel coordinate system,
minimizing the rms row and column differences between the astrometric and $r$
stars.  If this is the first fitting iteration, outliers are
rejected.  Smoothing splines, using three piecewise polynomials per frame,
are then fitted to the row and column residuals as a function of row (time).
Outliers are rejected, and the fitting process is iterated two more times.
Finally, the remaining residuals for all matched pairs are fitted, using a
least-squares fit to a cubic polynomial, as a function of column to
remove the optical distortions.
The resultant solution is used to transform the stars detected on the
astrometric CCD to the pixel coordinate system of the $r$ CCD.
If a star was detected
on both the leading and trailing astrometric CCDs within a column, then its
transformed
$r$ coordinates are averaged and the mean position is used.
These transformed stars
will then be matched to the Tycho-2 stars to yield a transformation from
$r$ pixel coordinates to great circle CMP coordinates.

\begin{figure}
\plotone{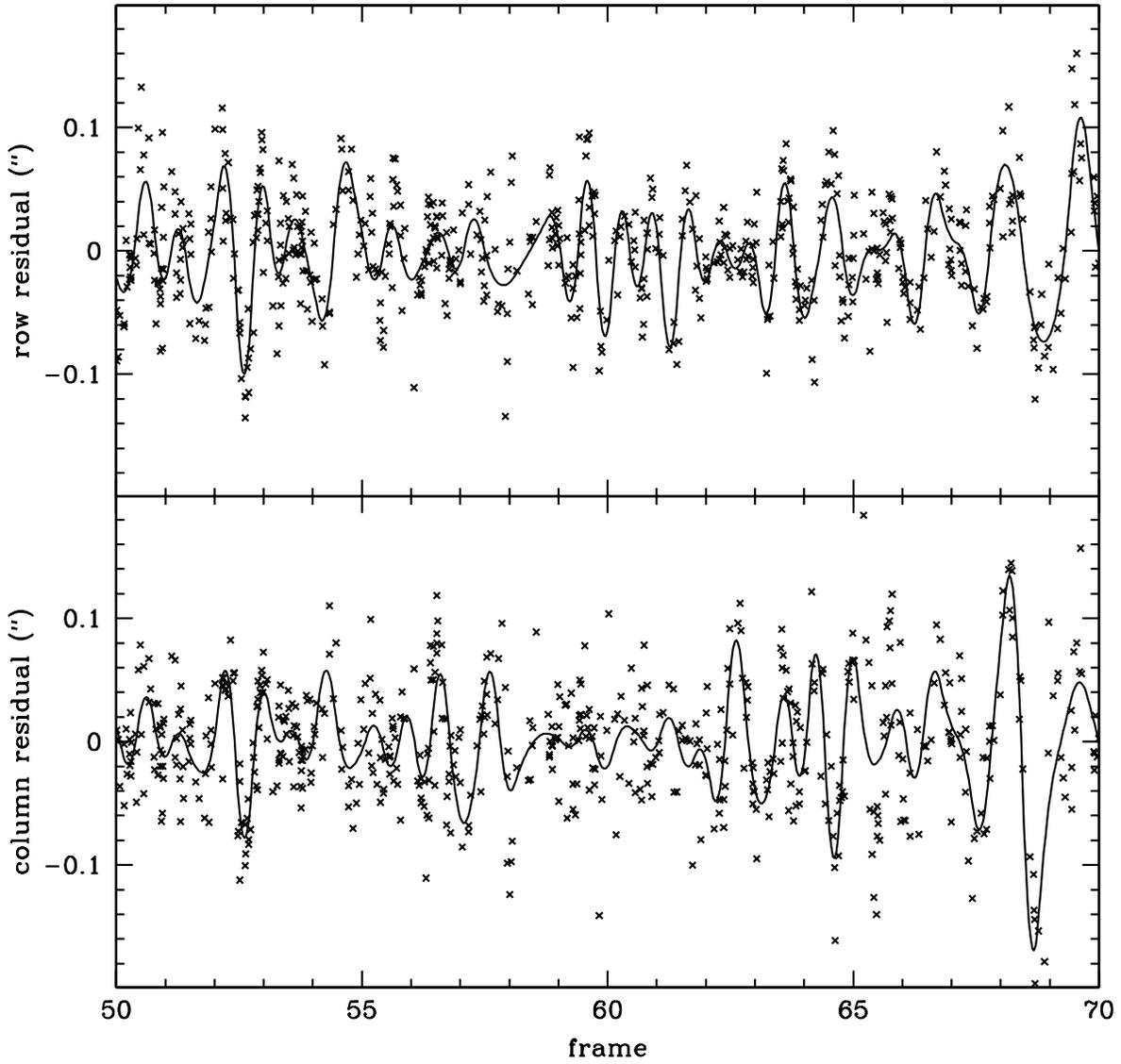}
\caption{Typical smoothing spline fits for the transformation from an
astrometric CCD to an $r$ CCD.  The 20 frames displayed is approximately
12 minutes of data.}
\label{fig-astrometricSplines}
\end{figure}

Figure~\ref{fig-astrometricSplines} shows a typical smoothing spline fit to the
row and column residuals for an astrometric CCD to $r$ CCD match.  The
difficulty in doing astrometry with 11 second exposures in drift scan mode
is evident.  Peak-to-peak excursions are 100 -- 300 mas, with time scales of
order half a frame (18 seconds).  These most likely are due to anomalous
refraction (see Section~\ref{sec-anomalous}).
The scan used for
the plot has a higher number of pairs per frame than most scans, with half
that number more typical.  While no all-sky astrometric catalog of sufficient
accuracy is dense enough to follow the atmospheric
excursions on these time scales, there are enough overlap
stars between the astrometric and $r$ CCDs to allow most of the systematics
to be removed.  There is a tendency for the peaks of the excursions to be
underfitted.  Interpolating splines were experimented with in place of the
smoothing splines, but the final astrometry was not improved.
The rms residual in either row or column for the transformation
from an astrometric CCD to an $r$ CCD is typically 30 -- 50 mas.

The reduction now proceeds similarly whether reducing against UCAC or Tycho-2.
The following focal plane model is used, relating frame row ($x$)
and column ($y$) to PCP coordinates (the small angle
approximation is assumed for all rotations).
\begin{eqnarray}
x' & = & x + g_0 + g_1 y + g_2 y^2 + g_3 y^3\\
y' & = & y + h_0 + h_1 y + h_2 y^2 + h_3 y^3\\
\mu_{\ast,PCP} & = & \mu_{B,CMP} + T_{\mu} x' + s [ X_{CCD} - X_B
     + (Y_{CCD} - Y_B) \theta_{CAM} \\
\nonumber
&&   + 0.024 f_{CCD} (y' - 1024) (\theta_{CAM} + \theta_{CCD}) ] \\
\nu_{\ast,PCP} & = & \nu_{B,CMP} + T_{\nu} x' + s [ -(X_{CCD} - X_B)
     \theta_{CAM} + Y_{CCD} - Y_B \\
\nonumber
&&   + 0.024 f_{CCD} (y' - 1024) ]
\end{eqnarray}
$x'$ and $y'$ are frame row and column corrected for optical
distortions.  The telescope
tracking (along the CMP great circle) is modeled by specifying
the boresight pointing when row 0 of the frame was read ($\mu_{B,CMP}$ and
$\nu_{B,CMP}$, in arcseconds) and the tracking rate parallel ($T_{\mu}$,
in arcseconds pixel$^{-1}$) and perpendicular ($T_{\nu}$) to the great circle.
A perfectly tracked scan would have $\nu_{B,CMP} = 0$, $T_{\nu} = 0$, and
$T_{\mu}$ equal to the sidereal rate.  The location
of the telescope boresight in the focal plane is specified by the constant
parameters $X_B$ (always set to 0) and $Y_B$ (-22.74~mm for north strips,
+22.74~mm for south strips).  $s$ is the telescope scale
(in arcseconds mm$^{-1}$) and $\theta_{CAM}$ the rotation of the camera
from its correct value (in radians).  $X_{CCD}$ and $Y_{CCD}$ are the location
of the center of the CCD (column 1024) in the focal plane (in mm),
$\theta_{CCD}$ is the
rotation of the CCD with respect to the focal plane coordinate system,
$f_{CCD}$ is the relative scale of the CCD (with nominal value of 1), and the
size of the pixels is 0.024~mm.
$\mu_{\ast,PCP}$ and $\nu_{\ast,PCP}$ are the resultant PCP
coordinates of the star (in arcseconds).

The camera geometry (the telescope scale and the CCD focal plane locations,
rotations, relative scales, and optical distortion terms) is
calibrated once a month using a scan through a UCAC field.
(Before UCAC became available, calibration regions measured using the
Flagstaff Astrometric Scanning Transit
Telescope [\citealp[FASTT;][]{fastt,fastter}] were used.)
The most recent set of these nominal values is used as an initial
(and in some cases final) guess at those values for the current reduction.
The CCD terms are stable from month to month.

The focal plane model, initialized using the TCC
telescope tracking information and a recent calibration scan, is first used to
calculate a priori CMP coordinates for each observed star
(stars detected on the $r$ CCDs for
reductions against UCAC, and stars detected on the astrometric CCDs and
converted to the $r$ pixel coordinate systems for reductions against Tycho-2).
These a priori coordinates are used to match the observed stars with the
catalog stars (catalog stars whose catalog error in either
right ascension or declination exceeds 60 mas are not used, rejecting about
25\% of the stars).  For each matched pair,
the position of the catalog star is converted from CMP to
PCP using the TT at mid-exposure for the observed star and
the color of the star.  UCAC does not contain color information, so the
observed color (\ri) of the star is used for reductions against UCAC.
Tycho-2 stars saturate the photometric CCDs and therefore do not have SDSS
colors; thus the Tycho-2 catalog colors ($\bvt$) are used, converted
to \ri\ using the transformation
\begin{equation}
\ri = -0.186 + 0.681 (\bvt) - 0.534 (\bvt)^2 + 0.319 (\bvt)^3.
\end{equation}

A solution to the focal plane model is now derived for each frame, in order to
minimize the rms differences in $\mu_{PCP}$ and $\nu_{PCP}$
between the observed and catalog stars.  First, a separate linear
least-squares fit is performed for each set
of six frames observed at the same
time by the six $r$ CCDs.
The CCD terms (focal plane locations,
scales, rotations, and optical distortions) are held constant at
their nominal values, and only the six terms specifying the boresight tracking
($\mu_{B,CMP}$, $\nu_{B,CMP}$, $T_{\mu}$, and $T_{\nu}$), telescope scale
($s$), and camera rotation ($\theta_{CAM}$) are fitted.  For a given set of six
frames, all matched pairs on that set of frames
are used.  Additional matched pairs from the nearest adjacent frames are used
if required to yield a minimum number of matched pairs (30 for reductions
against UCAC, 15 for reductions against Tycho-2).
If an adjacent set of frames is used, then all matched pairs on that set of
frames are used.
For reductions against UCAC, there are typically 3 -- 10 catalog stars per
frame, so it is usually not necessary to use the adjacent frames.  For
reductions against Tycho-2, there is typically one catalog star every two
frames; thus it is generally necessary to use the adjacent four or more sets
of frames to achieve the minimum number of pairs.

If this is the first iteration, outliers are rejected.
There are systematics remaining in the residuals after the least-squares fit,
primarily due to anomalous refraction, varying with
typical time scales of 5 -- 20 frames (see Section~\ref{sec-anomalous}).
For reductions against UCAC, smoothing splines are
fitted to the $\mu$ and $\nu$ residuals for each $r$ CCD separately, with
breakpoints separated by five frames.  This is not possible for reductions
against Tycho-2, due to the paucity of catalog stars.
Rather, the average (over the entire scan) residual in $\mu$ and $\nu$ is
removed for each $r$ CCD separately.
In either case, this procedure effectively refits the CCD location in the
focal plane ($X_{CCD}$ and $Y_{CCD}$).
For both the UCAC and Tycho-2 case, outliers are rejected, and the process of
the least-squares fit, spline fits, and outlier rejection is iterated two more
times.

\begin{figure}
\plotone{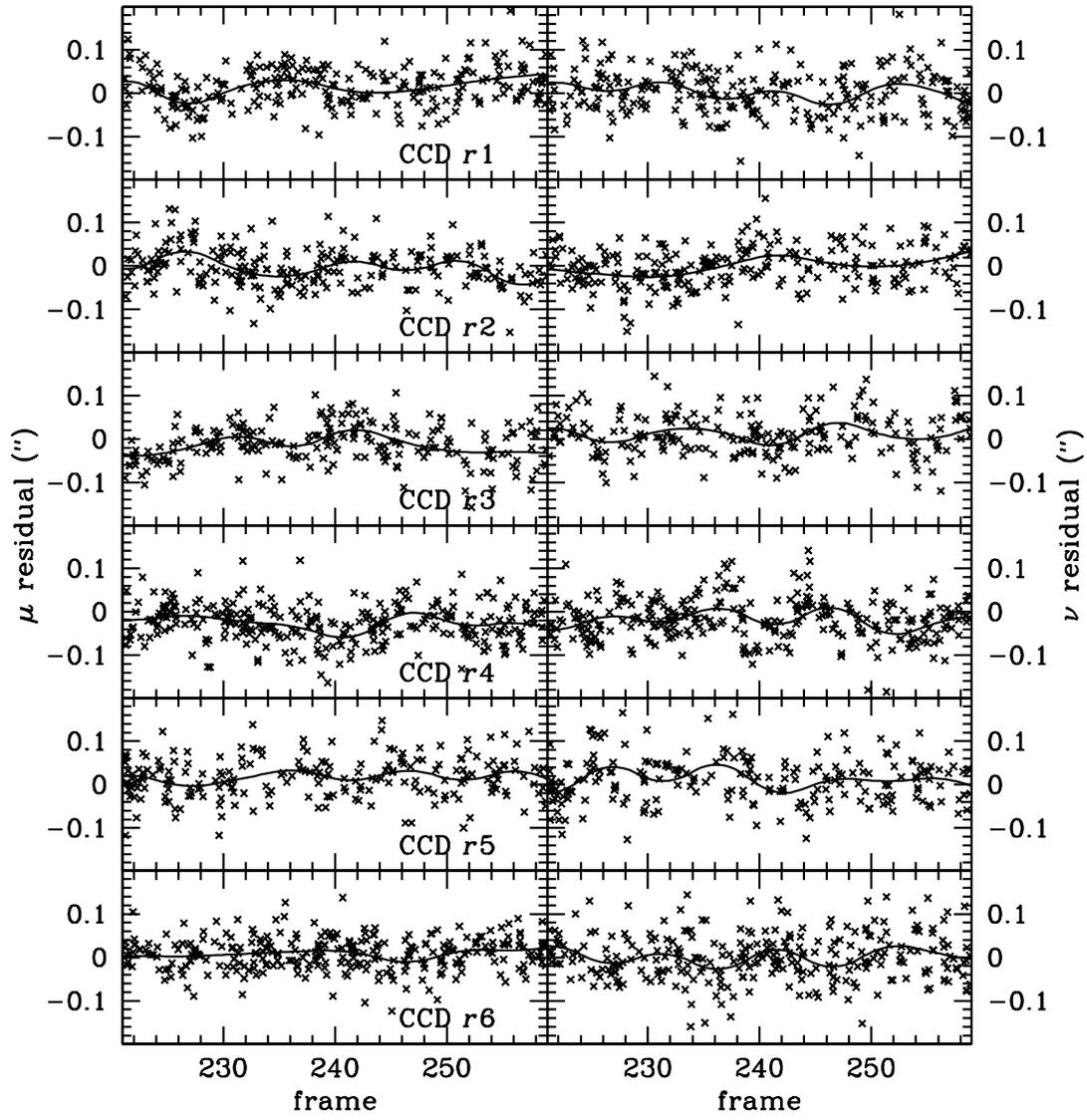}
\caption{Typical smoothing spline fits for the transformation from
$r$ CCDs to UCAC.}
\label{fig-ucacSplines}
\end{figure}

Figure~\ref{fig-ucacSplines} shows typical smoothing spline fits to the
$\mu$ and $\nu$ residuals for all six $r$ CCDs for part of a reduction against
UCAC (separate fits for each CCD).  Systematics in common to all six CCDs
(such as due to telescope jitter or any portion of anomalous refraction which
is correlated across the focal plane) have already been removed by the
running-means.  The remaining
systematics have typical amplitudes of 20 mas, with time scales of a few
frames or more.   There are occasional features with time scales closer to
a frame, which are more prevalent in worse seeing.  The spline fits remove
most of the systematics but not all.
More closely placed breakpoints have been experimented with, but this can
lead to fitting features that are not real.  The chosen breakpoint
separation of five frames is a good compromise between fitting most of the
real features without overfitting the data for typical seeing conditions.
These systematics cannot be
removed for reductions against Tycho-2, as there are too few catalog stars
detected per CCD to map the systematics.  This adds roughly 10 -- 20 mas
in quadrature to the rms errors for reductions against Tycho-2.
Similarly, the running-means typically smoothes over single frames for
reductions against UCAC, and five frames for reductions against Tycho-2.
Thus, any correlated systematics with time scales of order a
frame cannot be removed from reductions against Tycho-2.

For reductions against UCAC, the optical distortion terms
($g_i$, $h_i$) are then refitted by fitting the remaining residuals to a cubic
polynomial as a function of column.
One set of terms is calculated per CCD for the entire scan, not per frame.
This effectively refits
the CCD rotations and relative scales ($\theta_{CCD}$ and $f_{CCD}$ in the
focal plane model).
There are not enough matched pairs for a reduction against Tycho-2,
particularly for short scans, to refit these terms, so the nominal values
from the most recent calibration scan are used.  While these terms are
very stable over time, occasionally there is evidence of residual CCD
rotation or scale changes.

The solution is a transformation (nonlinear in row) from $r$ pixel
coordinates to PCP coordinates.  This needs to be converted to
the calibration equations of Section~\ref{sec-equations}, which convert from
pixel coordinates to CMP coordinates, one set of equations per
frame.  The optical distortion terms of the solution
are copied directly to the calibration equations.
For each frame, PCP coordinates are calculated from the
solution at the midpoint of each of the four sides of the frame.  These are
converted to CMP using the TT at mid-exposure and
assuming a star of zero color.  These four test points then determine the
affine transformation portion of the calibration equations.  The CMP
is similarly calculated for 11 test points all located in the middle of
the frame but spanning a range in color.  The $\mu_{CMP}$ and $\nu_{CMP}$
offsets from a star of zero color are then fitted as linear functions of color,
yielding the DCR terms for each frame.

\subsection{$u$, $g$, $i$, and $z$ Calibrations}

The $r$ calibration equations are applied to all the stars detected on the
$r$ CCDs,
generating CMP positions for these stars.  This creates a
catalog of secondary astrometric standards which is used to calibrate the
remaining photometric CCDs.  Each of the $u$, $g$, $i$, and $z$ CCDs is
calibrated separately.  The stars detected on a given CCD are matched against
the secondary catalog from the $r$ CCD in the same camera column.
There are typically 10 -- 40 matched pairs per frame (5 -- 20 on the $u$
CCDs).
For each matched pair, the position of the catalog star is converted from
CMP coordinates to PCP coordinates using the TT
at mid-exposure and the color (\ug\ for $u$, \gr\ for $g$, \ri\ for $i$ and
$z$) of the observed star.
For each frame all matched pairs on that frame are used to derive an affine
transformation from pixel coordinates to PCP coordinates,
minimizing the rms differences in $\mu_{PCP}$ and $\nu_{PCP}$
between the observed and catalog stars.
If there are fewer than 15 matched pairs on the frame, then the nearest
matched pairs on adjacent frames are used to guarantee that a minimum
of 15 pairs is used.
Outliers are rejected, and the process of
fitting and outlier rejection is iterated two more times.
The remaining residuals for all matched pairs are fitted using a least-squares
fit to a cubic polynomial as a function of column to remove the
optical distortions (that is, one polynomial is fitted per CCD for the entire
scan, not each frame).  Chromatic aberration in the optics can produce
systematics with color in the remaining
residuals.  This is significant only in $g$.  Thus, for $g$ only, an
empirically determined linear term with color (based on a fit to a single
scan, but which has proven to be stable from scan to scan) is removed
from the remaining residuals.
The resulting solution transforms pixel coordinates to PCP
coordinates.  It is converted to the calibration equations of
Section~\ref{sec-equations}, which convert from pixel coordinates to
CMP coordinates, using the same procedure used for the
$r$ CCDs.

\section{Results}
\label{sec-results}

This section quantifies the accuracy of the astrometric calibrations.
The calibrations are based on stars brighter than $20^{th}$ magnitude,
for which centroiding errors are negligible (except in the $u$ filter, where
the centroiding errors start to become significant by $20^{th}$
and contribute to the rms errors quoted here).
Thus, the accuracies quoted here are for the calibrations only, and reflect
the astrometric accuracy expected for stars brighter than $20^{th}$.
Centroiding errors (which are calculated by \frames) must also be considered
for fainter stars and all galaxies, and are discussed at the end of the
section. The results presented here are based on the entire SDSS data set to
be made public in early 2003 as SDSS Data Release 1 (DR1).
This includes 33 scans covering roughly 1200 square degrees of sky (not all
unique) reduced against UCAC, and 32 scans covering roughly
750 square degrees of sky reduced against Tycho-2.
The distribution of internal residuals within a single scan,
the distribution of differences in position for matched pairs between two runs,
and the distribution of differences in position for the SDSS matched
against the external catalogs used in this analysis are all well
characterized by Gaussians, and therefore by rms residuals and differences.
Thus all errors are quoted as rms residuals or differences in $\mu$ and $\nu$.

\subsection{r Astrometry}

\begin{figure}
\plotone{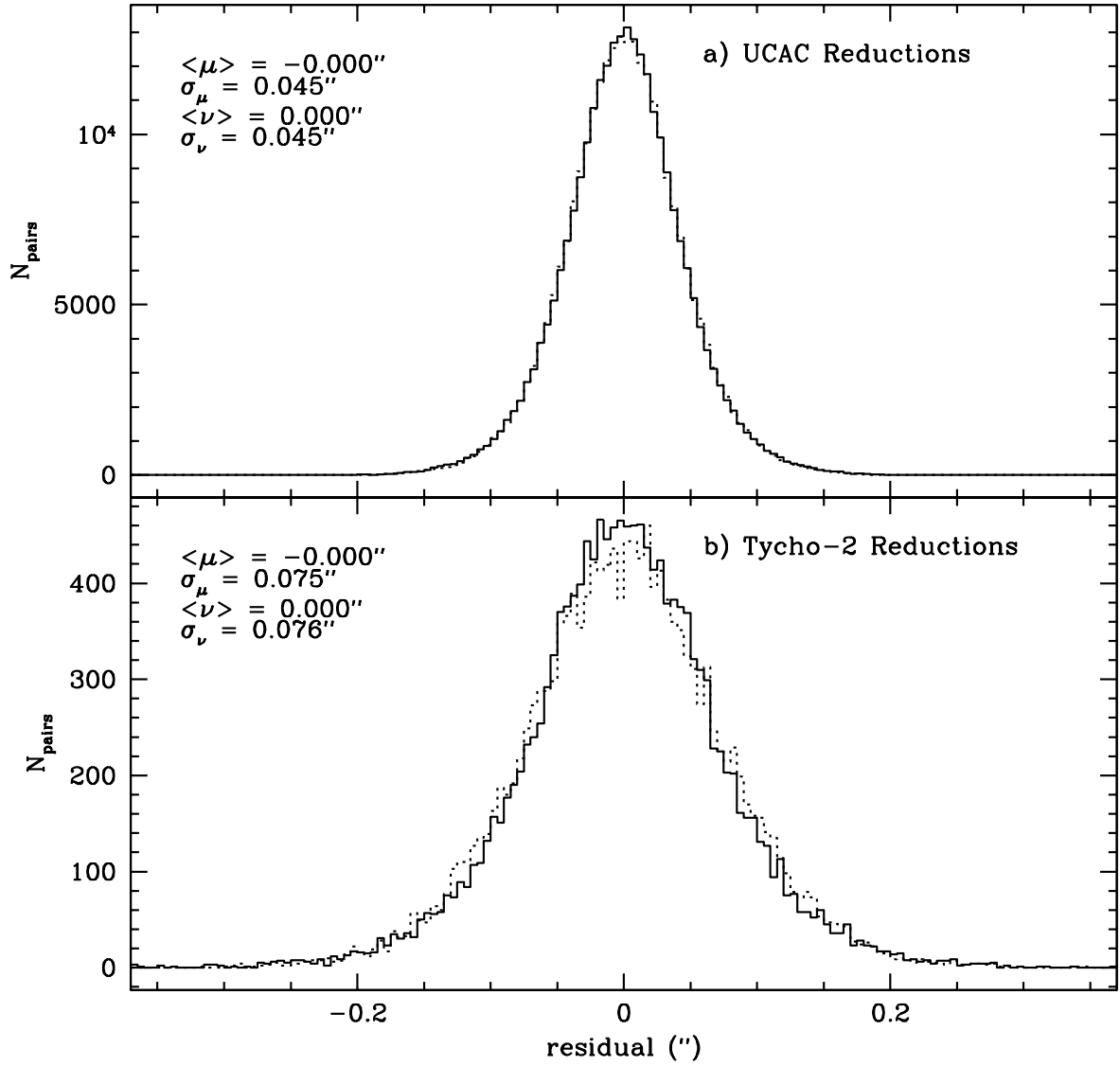}
\caption{Histograms of the residuals (SDSS position minus
catalog position) for scans reduced against UCAC (a) and scans reduced
against Tycho-2 (b).
The solid and dotted histograms are for the $\mu$ and $\nu$ residuals,
respectively.}
\label{fig-internalAbsDiff}
\end{figure}

The absolute accuracy of the $r$ astrometry is difficult to gauge, as there are
no astrometric catalogs as deep and accurate as the SDSS itself.
The primary internal measure of the precision of the astrometry
is the distribution of residuals (measured
position of the SDSS object minus the catalog position) for
matched pairs (after outlier rejection) on the $r$ CCDs.
Figure~\ref{fig-internalAbsDiff} shows histograms of these residuals for all
matched pairs for reductions against UCAC and Tycho-2 separately.
The rms residuals per coordinate are 45 mas and 75 mas for reductions
against UCAC and Tycho-2, respectively.

\begin{figure}
\plotone{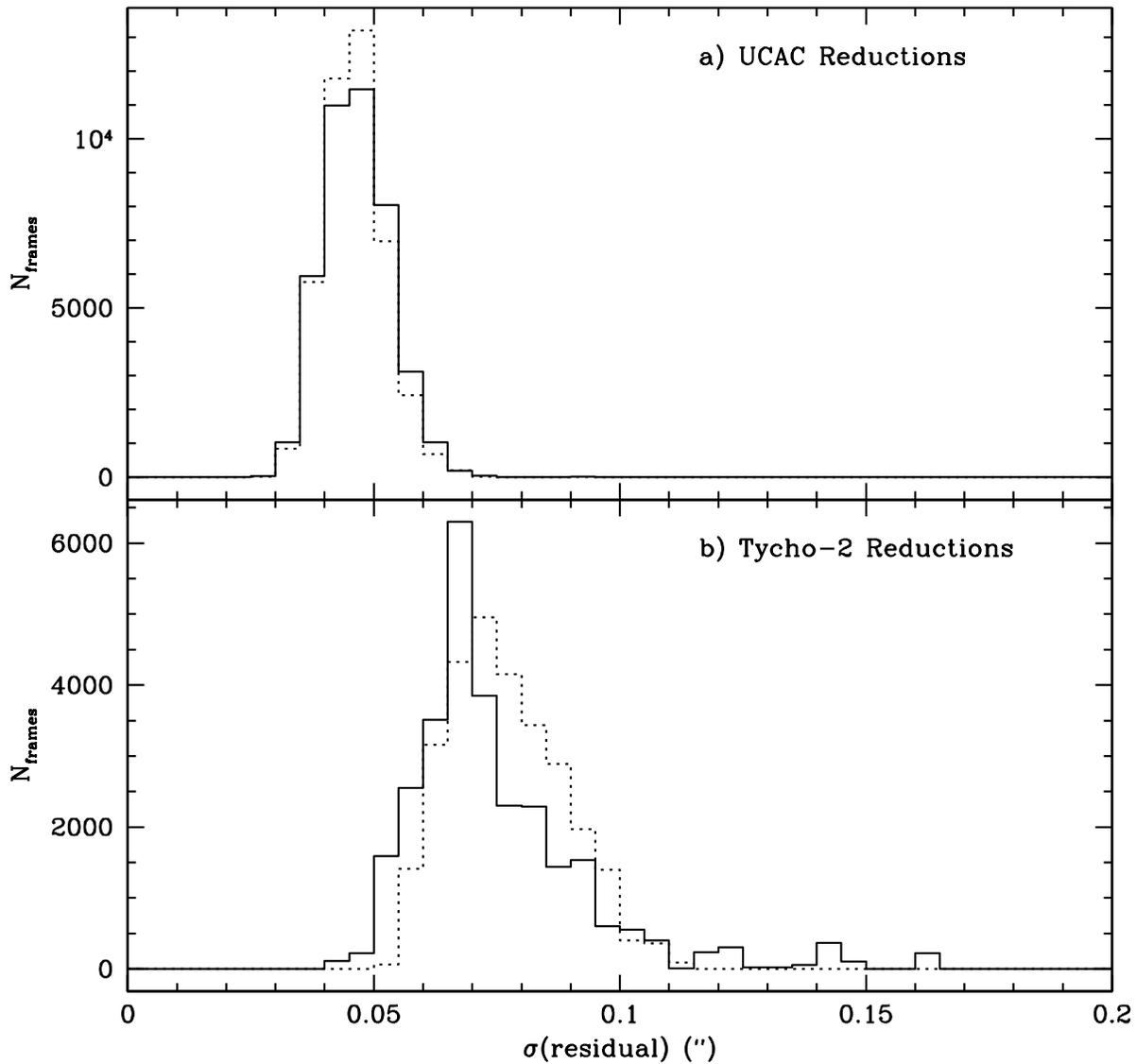}
\caption{Histograms of the rms residuals (SDSS position minus
catalog position) for each frame in scans reduced against UCAC (a) and
scans reduced against Tycho-2 (b).  The rms residuals for each frame are
calculated using a minimum of the nearest 100 matched pairs.
The solid and dotted histograms are for the $\mu$ and $\nu$ rms residuals,
respectively.}
\label{fig-internalAbsRms}
\end{figure}

\begin{figure}
\plotone{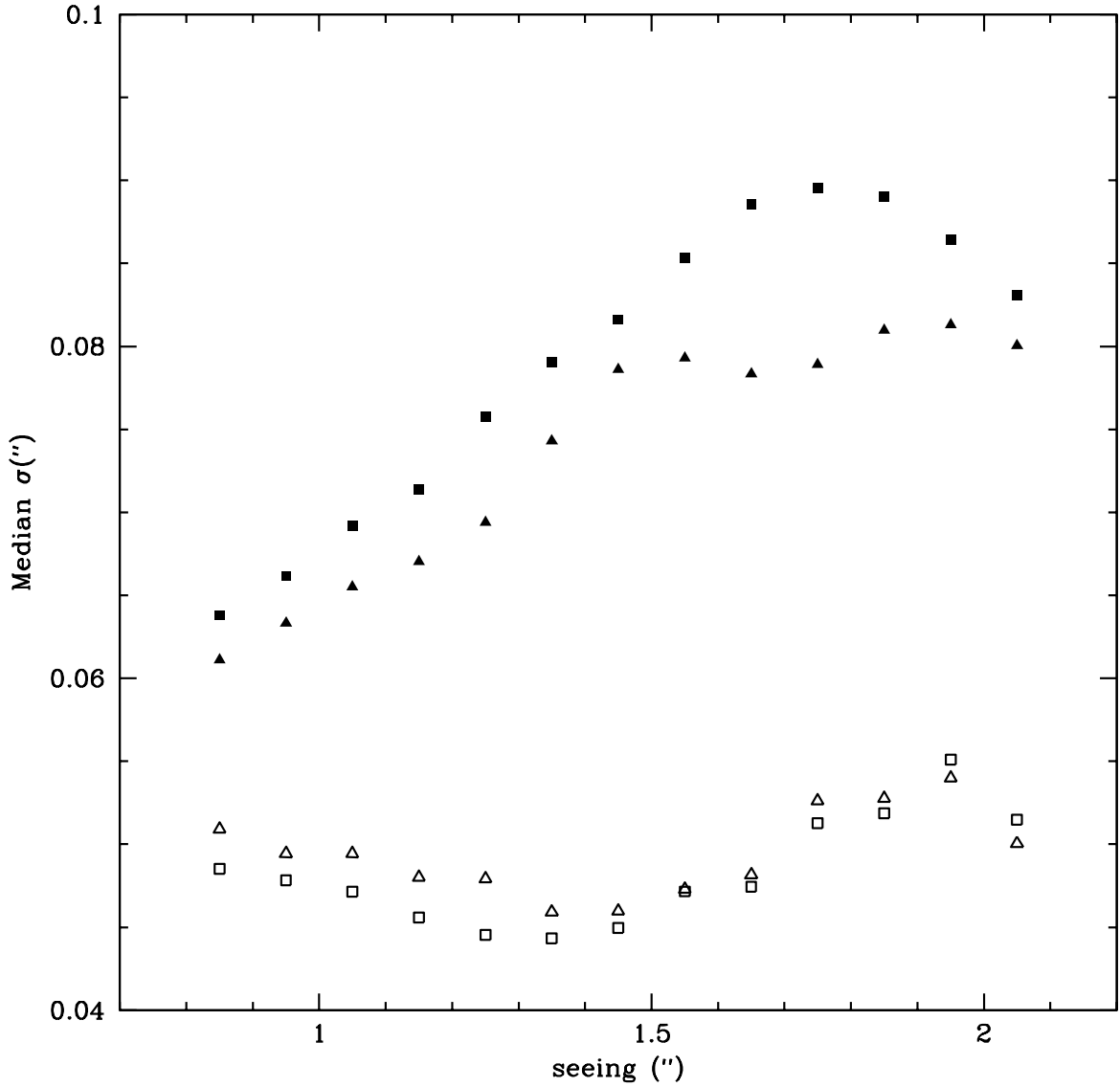}
\caption{Median rms residual per frame binned by seeing for the $r$ CCDs.
Triangles are for the
$\mu$ residuals, squares the $\nu$ residuals.  Open symbols are for reductions
against UCAC, filled symbols are for reductions against Tycho-2.}
\label{fig-absSeeing}
\end{figure}

\begin{figure}
\plotone{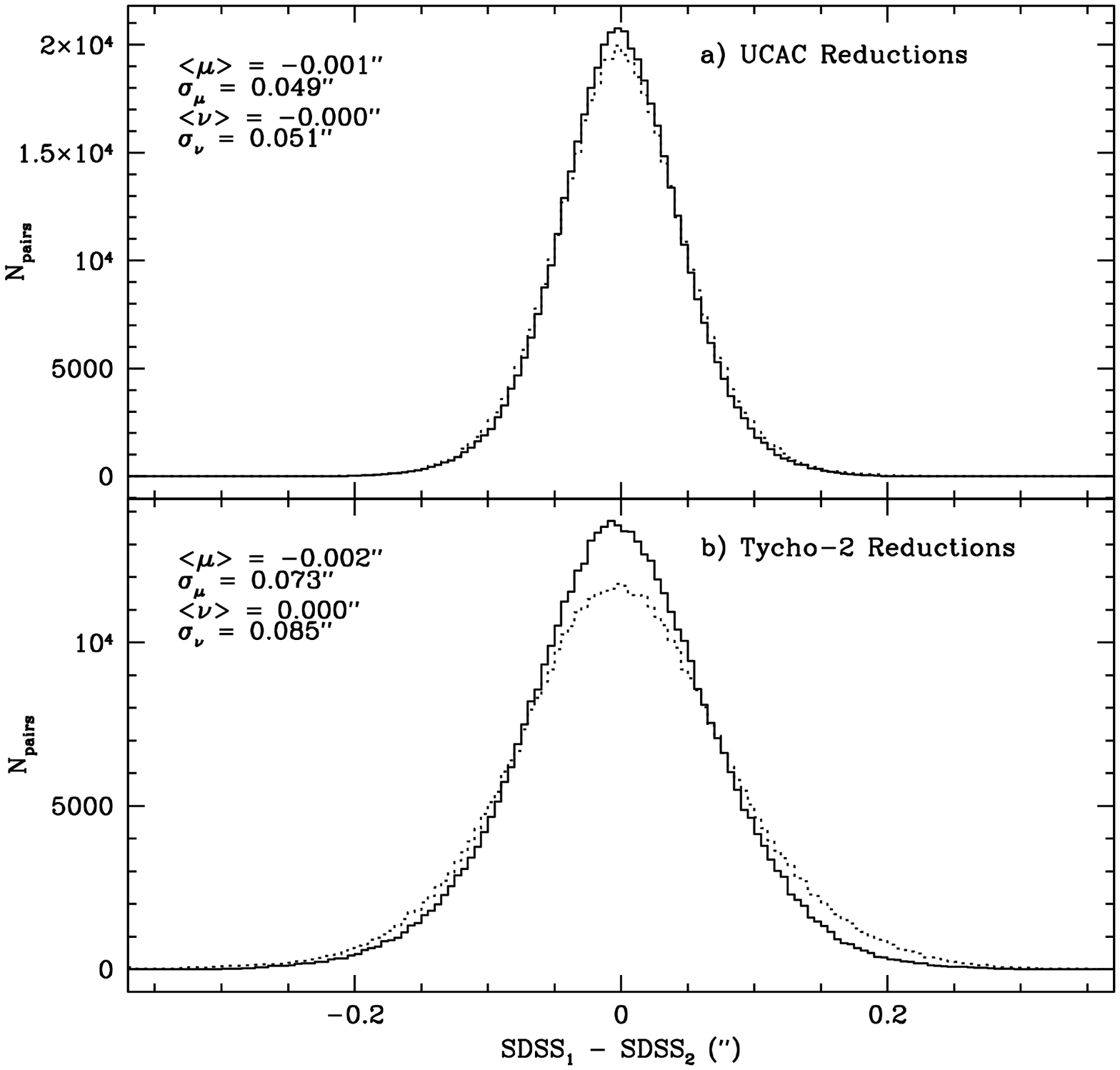}
\caption{Histograms of the position differences for matched pairs
on overlapping scans.
Results from reductions against UCAC are shown in panel a,
those for reductions against Tycho-2 in panel b.
The solid and dotted histograms are for the $\mu$ and $\nu$ differences,
respectively.}
\label{fig-runToRunDiff}
\end{figure}

The precision, of course, can vary from scan to scan, and during individual
scans.  For each scan the rms residuals (around 0) in
$\mu$ and $\nu$ are calculated separately for each frame, for each $r$
CCD separately.  To assure good statistics,
a smoothing window is used, using enough adjacent frames to give a minimum
of 100 matched pairs per frame.
These rms residuals in $\mu$ and $\nu$ for each frame are recorded with
the calibration equations for that frame, and serve as the best estimate
of the systematic errors for that frame.
Figure~\ref{fig-internalAbsRms} shows
histograms of these rms residuals for all frames in DR1.
All the distributions peak near the rms values
indicated in Figure~\ref{fig-internalAbsDiff}.
For reductions against UCAC almost all frames have rms residuals of less than
60 mas.  The distributions for reductions against Tycho-2 are broader,
with most frames having rms residuals of less than 100 mas.
The dependence on seeing is shown in Figure~\ref{fig-absSeeing}, which plots
the median rms residual per frame binned by seeing (as measured by \psp).
The reductions against
Tycho-2 are considerably more sensitive to seeing than the reductions against
UCAC, demonstrating that the longer integration times on the $r$ chips
(relative to the astrometric chips), coupled with the greater star density of
the UCAC catalog, are sufficient to remove most of the atmospheric effects.

\begin{figure}
\plotone{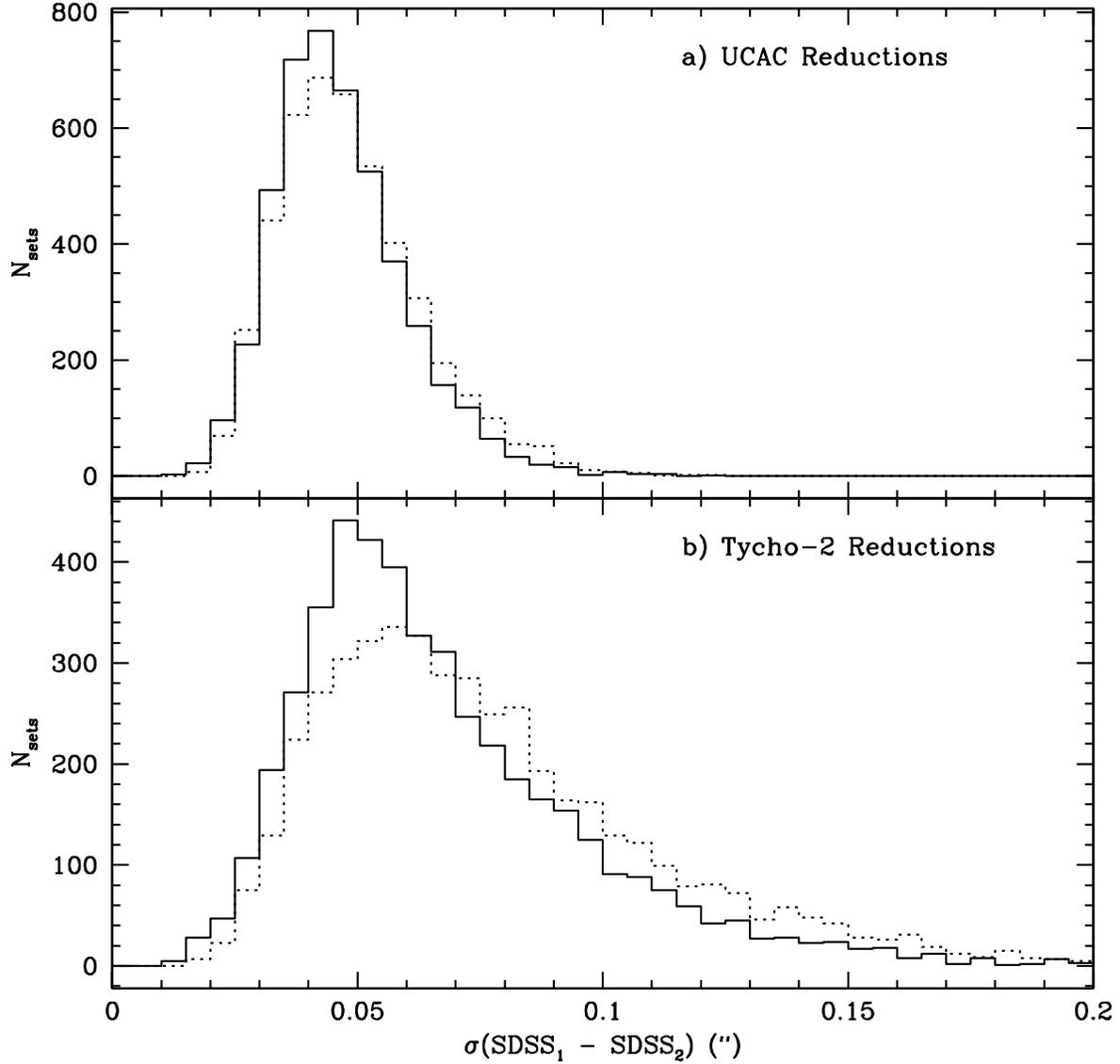}
\caption{Histograms of the rms position differences for sets of
100 matched pairs between overlapping scans.  For each pair of overlapping
scans and for each $r$ CCD the matched pairs are sorted by $\mu$ and
binned into groups of 100 matched pairs, and the rms differences in $\mu$ and
$\nu$ are calculated for each set of 100 matched pairs.
Results from reductions against UCAC are shown in panel a,
those for reductions against Tycho-2 in panel b.
The solid and dotted histograms are for the $\mu$ and $\nu$ rms differences,
respectively.}
\label{fig-runToRunRms}
\end{figure}

Some parts of the sky have been scanned multiple times.
Stars in nine pairs of overlapping scans (including some non-DR1
scans), reduced both against UCAC and Tycho-2, have been matched.  The
distribution of position differences for all matched pairs is shown in
Figure~\ref{fig-runToRunDiff}.
The rms differences are about 50 mas in both $\mu$ and $\nu$ for reductions
against UCAC, and 73 mas in $\mu$ and 85 mas in $\nu$ for reductions against
Tycho-2.  These are smaller than one might expect from simply adding the
internal errors in quadrature.
Errors in UCAC and Tycho-2 catalog positions, which propagate
to SDSS positions calibrated using those catalogs, are correlated among the
repeat scans and thus are partially removed when differences in the SDSS
positions are computed.
Variation of astrometric precision within and between scans
can be tested by sorting the matched pairs by $\mu$ and binning them
into groups of 100, separately for each $r$ CCD.
The rms differences in $\mu$ and $\nu$ are then calculated for each set of
100 matched pairs.  Figure~\ref{fig-runToRunRms}
shows histograms of these rms differences.
For reductions against UCAC, the
distributions peak around 45 mas, with most of the
data having rms differences of less than 70 mas.  For
reductions against Tycho-2 the distributions are fairly broad, with typical
rms differences of 60 mas, and most of the data
having rms differences of less than 130 mas.

\begin{figure}
\plotone{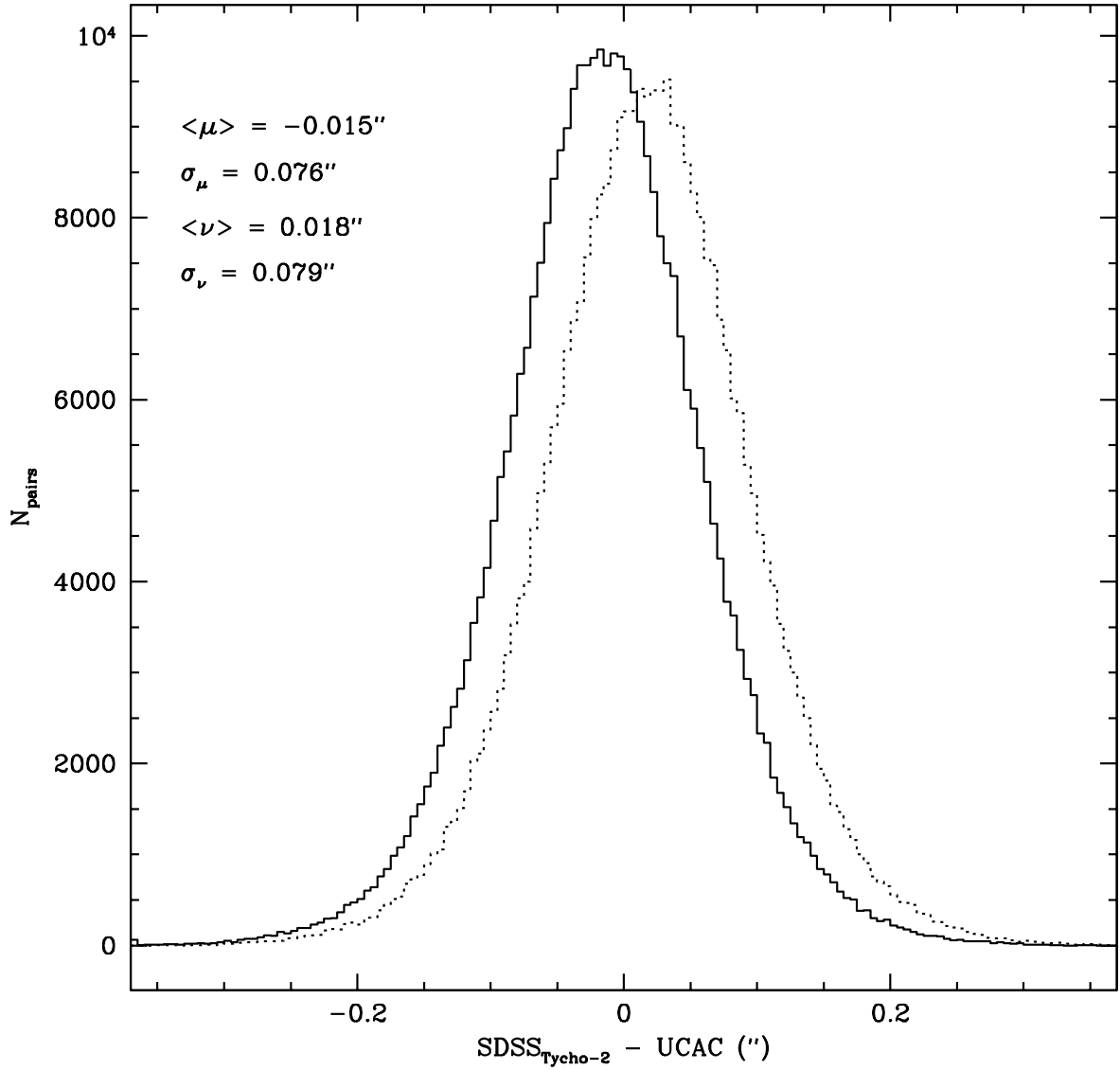}
\caption{Histograms of the position differences for matched
pairs for reductions against Tycho-2 matched against UCAC.
The solid and dotted histograms are for the $\mu$ and $\nu$ differences,
respectively.}
\label{fig-tycho2-ucac-diff}
\end{figure}

\begin{figure}
\plotone{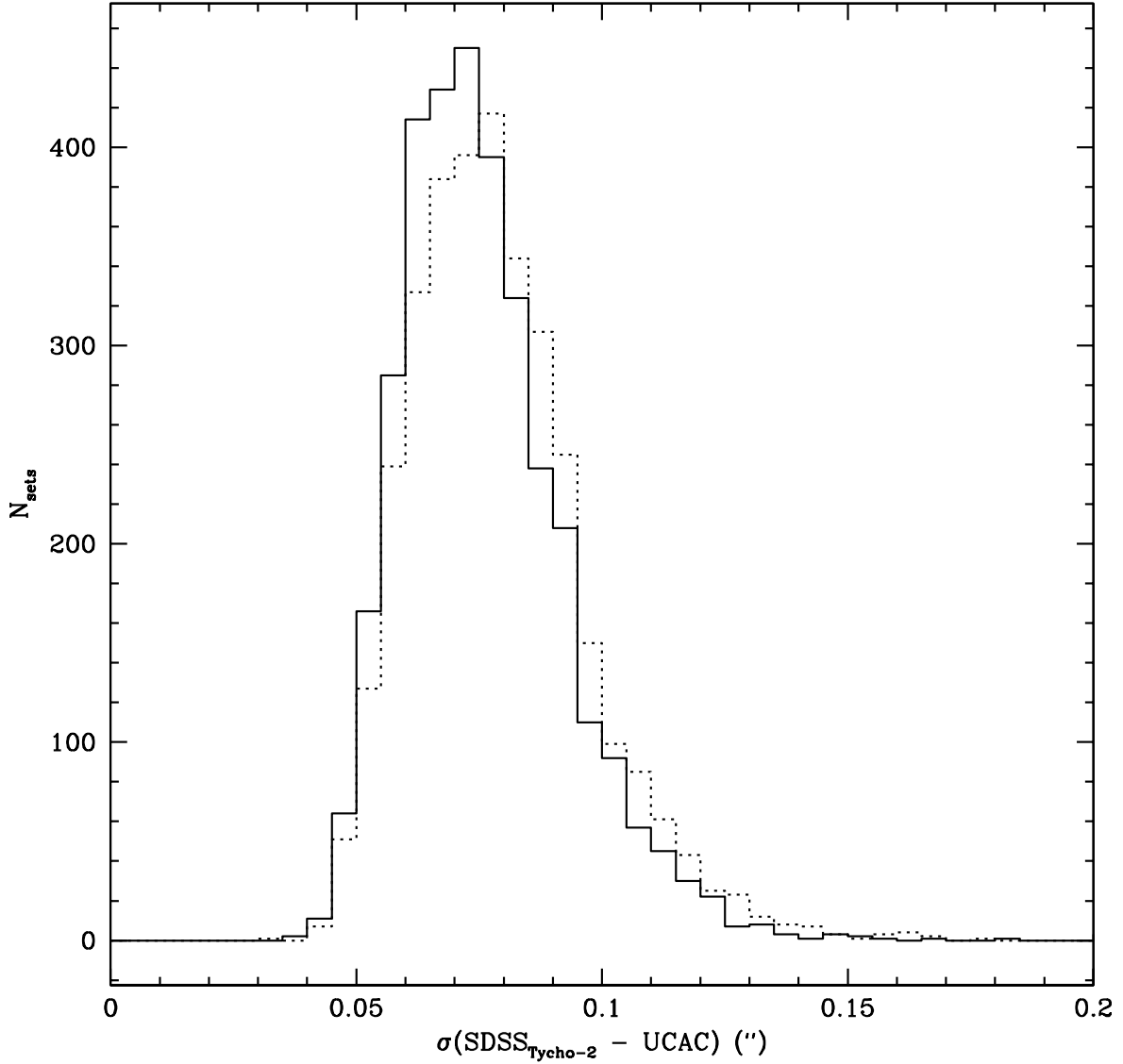}
\caption{Histograms of the rms position differences for sets of
100 matched
pairs, binned in $\mu$, for reductions against Tycho-2 matched against UCAC.
For each scan and for each $r$ CCD the matched pairs are sorted by $\mu$ and
binned into groups of 100 matched pairs, and the rms differences in $\mu$ and
$\nu$ are calculated for each set of 100 matched pairs.
The solid and dotted histograms are for the $\mu$ and $\nu$ rms differences,
respectively.}
\label{fig-tycho2-ucac-rms}
\end{figure}

An external measure of the accuracy of the $r$ astrometry for data reduced
against Tycho-2 may be obtained by re-reducing those scans which were reduced
against UCAC against Tycho-2, and matching the resultant star positions
against UCAC.  Histograms of the position
differences for all matched pairs are shown in
Figure~\ref{fig-tycho2-ucac-diff}.  The rms differences are 76 mas in $\mu$
and 79 mas in $\nu$, consistent with the internal rms residuals
(Figure~\ref{fig-internalAbsDiff}).  There are mean offsets of -15 mas in
$\mu$ and 18 mas in $\nu$.  These are likely due to uncorrected magnitude
terms in star positions on the SDSS astrometric CCDs, due to charge transfer
efficiency (CTE) effects (seen by matching the leading and trailing astrometric
CCDs in a given camera column), though residual
magnitude terms in the preliminary UCAC catalog used may also contribute.
The astrometric CCD magnitude terms are not currently corrected as 1) the
necessary test data are lacking, and 2) UCAC coverage should be
full-sky within a year, allowing all SDSS data to be recalibrated against
UCAC, bypassing the astrometric CCDs.
Again, variation of astrometric accuracy within and between scans
can be tested by sorting the matched pairs by $\mu$ and binning them
into groups of 100, separately for each $r$ CCD.
The rms differences in $\mu$ and $\nu$ are
then calculated for each set of 100 matched pairs.
Histograms of these rms differences are plotted in
Figure~\ref{fig-tycho2-ucac-rms}.  The distributions are similar to the
histograms for the rms internal residuals (Figure~\ref{fig-internalAbsRms}),
peaking at 70 mas, though with a
somewhat larger fraction of the data with rms differences exceeding 100 mas.

\begin{figure}
\plotone{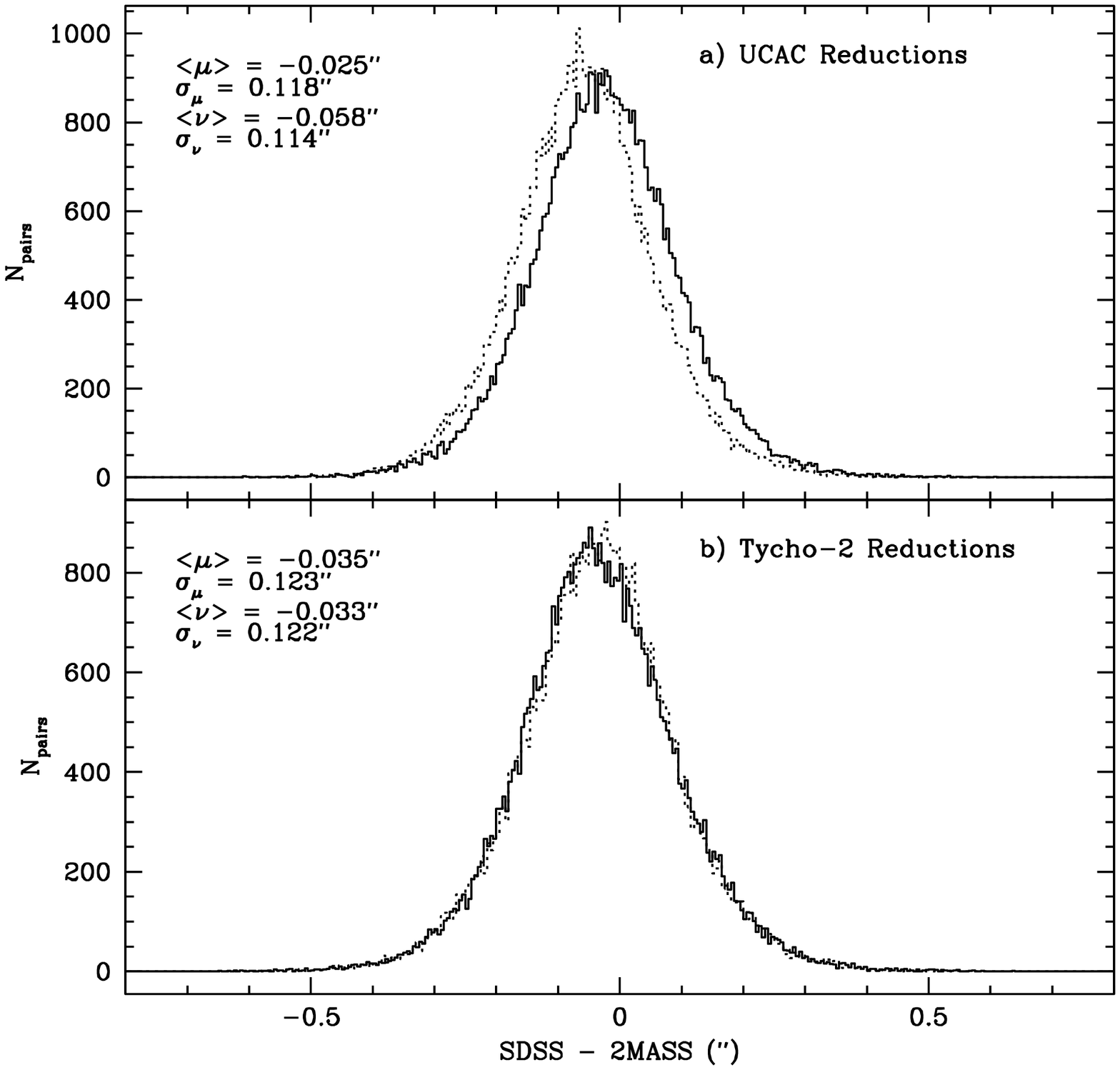}
\caption{Histograms of the position differences for matched pairs
between the SDSS and 2MASS for nine scans.
Results from reductions against UCAC are shown in panel a,
those for reductions against Tycho-2 in panel b.
The solid and dotted histograms are for the $\mu$ and $\nu$ differences,
respectively.}
\label{fig-sdss2mass}
\end{figure}

The external catalog which most closely matches the SDSS in terms of depth,
accuracy, and sky coverage is the Two Micron All Sky Survey
\citep[2MASS,][]{2mass}.  Typical position errors for the 2MASS Second
Incremental Release
are quoted as about 140 mas, though those are probably overstated and a truer
measure of the rms errors may be closer to 110 mas \citep{2mass-2ir}.
Figure~\ref{fig-sdss2mass} shows histograms of the differences between the
SDSS and 2MASS positions for stars in nine scans.  2MASS sources were
limited to non-confused, non-extended sources detected in all three bands
with $J < 15.8$, $H < 15.1$, and $K < 14.3$, and with the major axis of the
position error ellipse less than 175 mas.
The rms differences are 118 mas in $\mu$ and 114 mas in $\nu$
for SDSS data reduced against UCAC, and 123 mas in $\mu$ and 122 mas in $\nu$
for SDSS data reduced against Tycho-2.  These values are roughly what
one would expect by combining the rms errors in the two catalogs
in quadrature.
There are mean offsets of -25 mas in $\mu$ and
-58 mas in $\nu$ for SDSS reductions against UCAC, and -35 mas in $\mu$ and
-33 mas in $\nu$ for SDSS reductions against Tycho-2.  While the offsets in
$\mu$ could be explained by systematic errors in the SDSS reductions,
the $\nu$ offsets cannot.  Similar offsets in $\nu$ are seen when
matching UCAC against 2MASS in the same region of sky.  A known bias exists
in the 2MASS Second Incremental Release reductions whereby the positions of
bright stars ($K \lesssim 9$), which are processed through the $R1$ path of
the 2MASS pipeline, can be offset from those of fainter stars, which are
processed through the $R2-R1$ path of the 2MASS pipeline; this bias
will be addressed in the next 2MASS data release (H. L. McCallon 2001,
private communication).
This bias is likely the largest contributor to the $\nu$ offsets between
the SDSS and 2MASS, and may also contribute to the $\mu$ offsets.
(An earlier comparison between SDSS and 2MASS astrometry, based on a smaller
data set, is presented by \citeauthor{finlator} [2000].)

\begin{figure}
\plotone{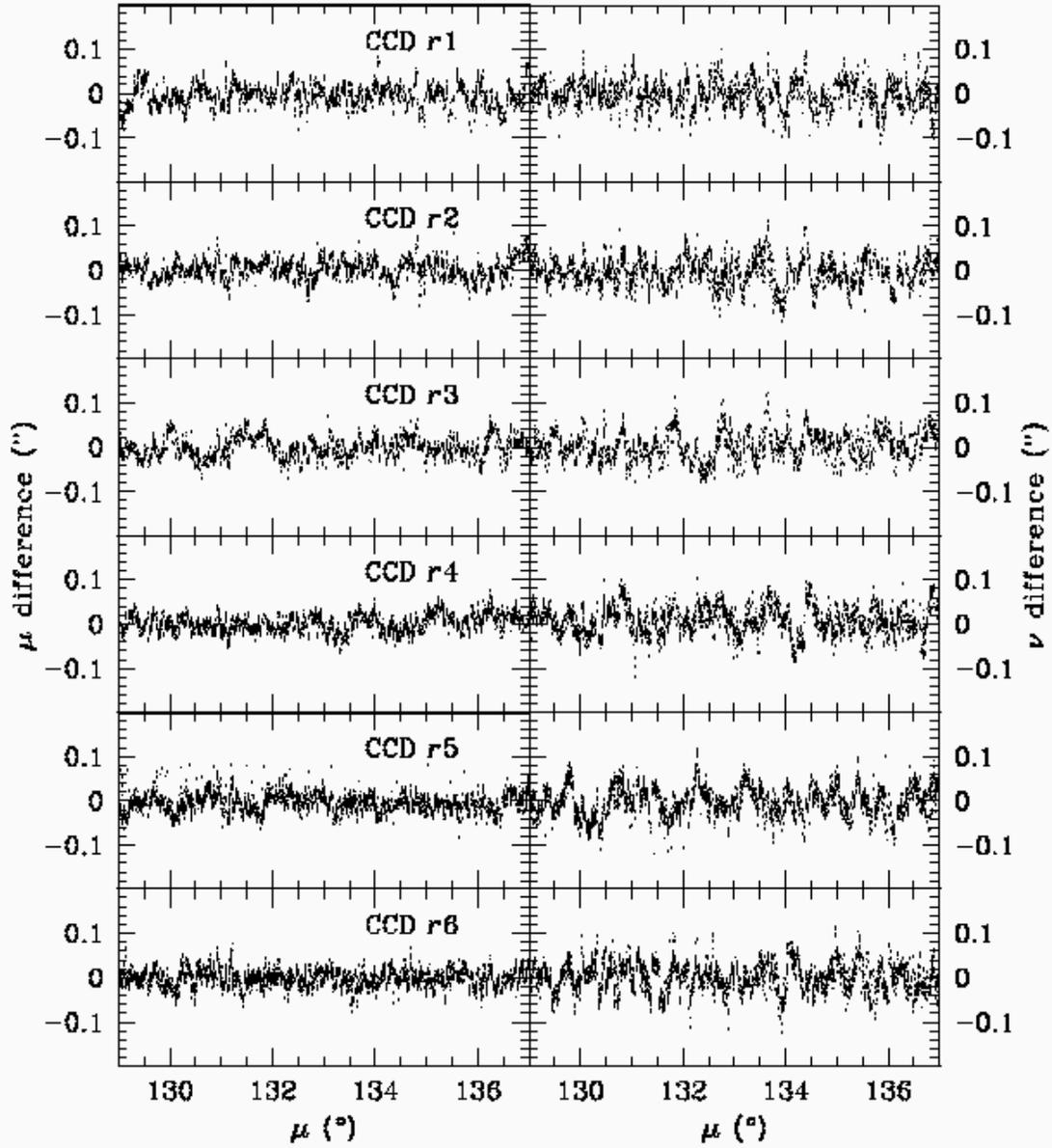}
\caption{Differences in position between matched stars in two
overlapping
scans reduced against UCAC plotted against $\mu$ (time).  The $\mu$ and $\nu$
differences are plotted separately for each $r$ CCD.}
\label{fig-run-to-run-time-ucac}
\end{figure}

\begin{figure}
\plotone{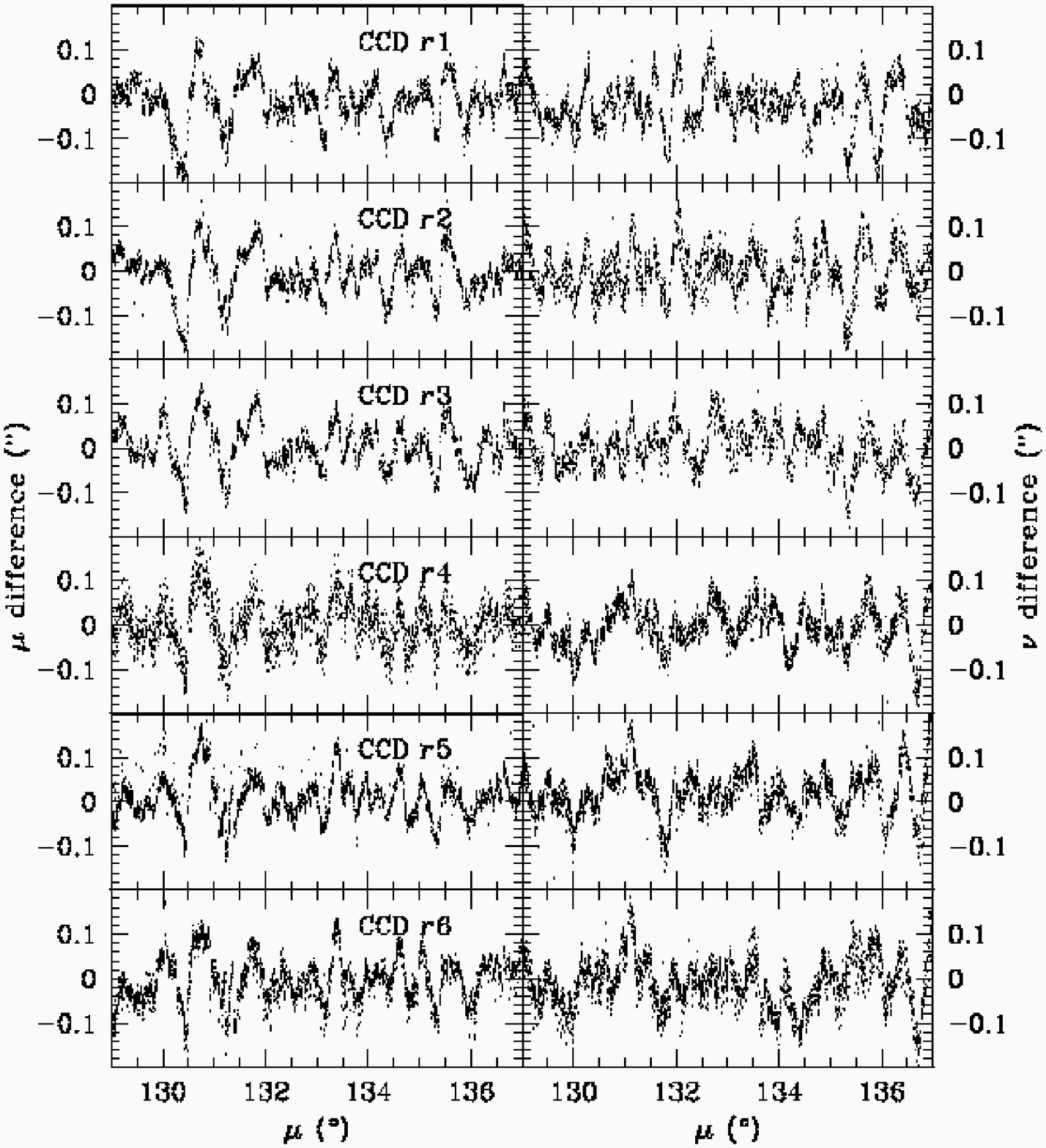}
\caption{Same as Figure~\ref{fig-run-to-run-time-ucac}, but
these data have been reduced against Tycho-2.}
\label{fig-run-to-run-time-tycho2}
\end{figure}

It should be emphasized that the values of the rms residuals and differences
quoted characterize the
distribution of {\em systematic} errors.  The analysis is based on stars
with $r < 20$, for which random centroiding errors are small compared to the
systematic errors.
Figure~\ref{fig-run-to-run-time-ucac} displays the differences in $\mu$ and
$\nu$ between matched
stars in two overlapping scans reduced against UCAC plotted against
$\mu$ (time).  The $\mu$ and $\nu$ differences
are plotted separately for each $r$ CCD.
Figure~\ref{fig-run-to-run-time-tycho2} shows the differences for matches for
the same portion of the same overlapping scans, but reduced against Tycho-2.
The differences vary systematically with time, on time scales of order one to a
few minutes.  While the differences are mostly uncorrelated between CCDs for
the reductions against UCAC, they are strongly correlated for the reductions
against Tycho-2.  There are simply too few Tycho-2 stars to adequately follow
the tracking errors and that portion of anomalous refraction
which is correlated across the camera on short time scales.  These plots are
typical, and display well the nature of the dominant systematics in
the astrometry.

SDSS astrometry is subject to any systematics present in the
primary astrometric reference catalog.  At the current epoch, systematics
in Tycho-2 are expected to be less than a few mas, and as such
do not degrade the overall accuracy of SDSS astrometry.  Uncorrected
magnitude terms in the star positions on the astrometric CCDs, used to
transfer the Tycho-2 stars to the photometric CCDs,
introduce systematic offsets between the Tycho-2 frame and SDSS data reduced
against Tycho-2 of order 20 mas.
A preliminary, internal version of UCAC (r07) is currently being used
for SDSS astrometric reductions.  This version of the catalog
contains some uncorrected magnitude terms, such that the brighter
stars, which are used to calibrate UCAC against Tycho-2, may be
systematically offset from the fainter stars, which are used to
calibrate the SDSS.  Such offsets are thought to be of order 10 to 20
mas, mainly caused by residual CTE effects in the CCD used for the 
UCAC observations.
In addition there will be zeropoint errors due to the fact
that a given UCAC frame can only be linked with a certain accuracy
to the Tycho-2 stars, depending on the number of reference stars
available in that area of the sky.  This will cause systematic
errors of up to 30 mas on the 0.5 to 1 degree scale.  
Since UCAC uses only a single bandpass, colors for the
stars were not available and DCR corrections could not be
applied.  However, the narrow bandpass used causes only 5 to 10 mas 
of systematic errors as a function of spectral type and zenith distance.  In
sum, total systematic errors in UCAC are thought not to exceed 30 mas
(N. Zacharias 2001, private communication).  As newer versions of UCAC
with smaller systematic errors become available, the SDSS
astrometry will be recalibrated.


\subsection{$u$, $g$, $i$, and $z$ Astrometry}

\begin{figure}
\plotone{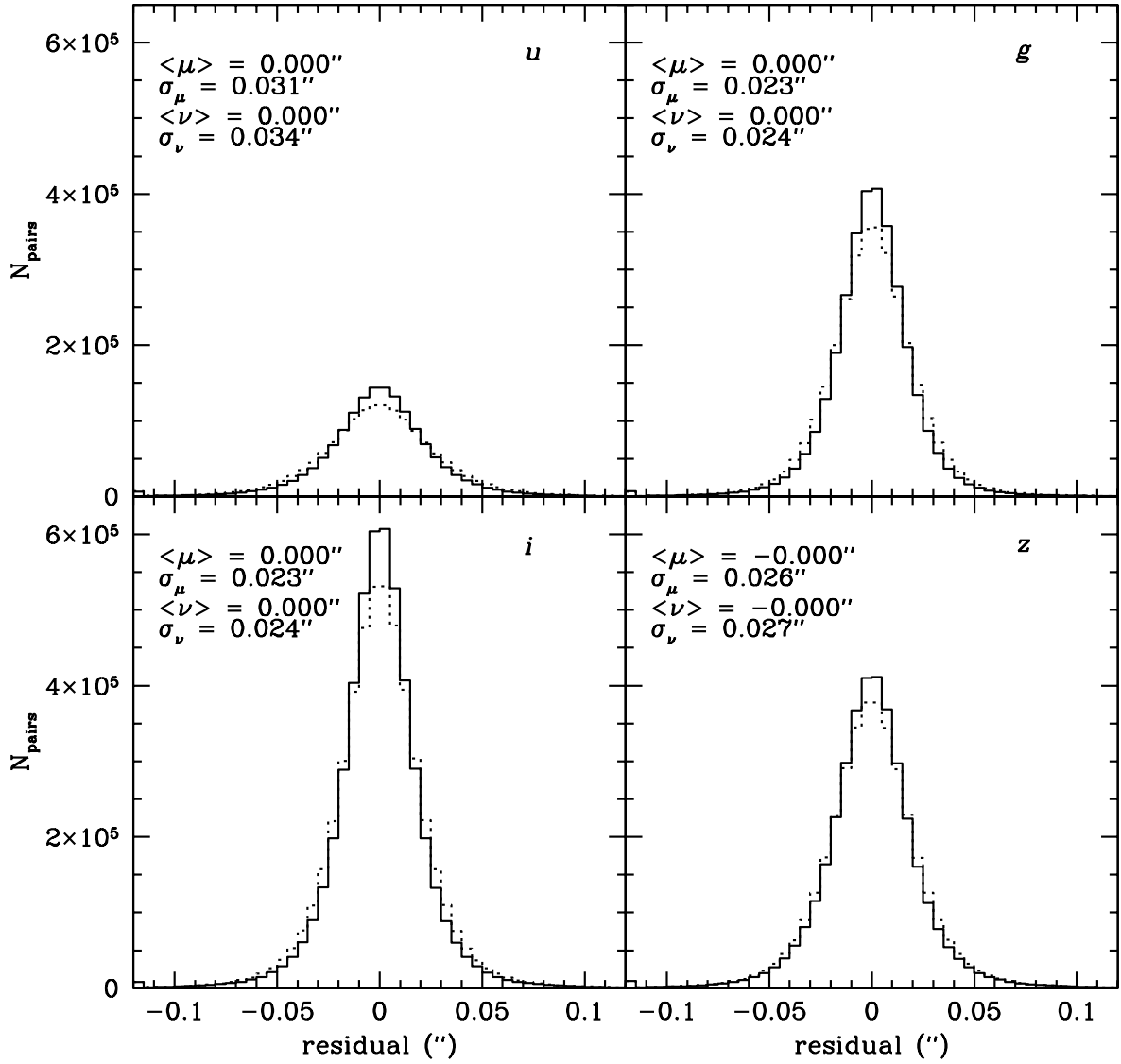}
\caption{Histograms of the residuals (SDSS position in the
specified filter minus SDSS position in the $r$ filter) for all DR1 scans.
The solid and dotted histograms are for the $\mu$ and $\nu$ residuals,
respectively.}
\label{fig-internalRelDiff}
\end{figure}

\begin{figure}
\plotone{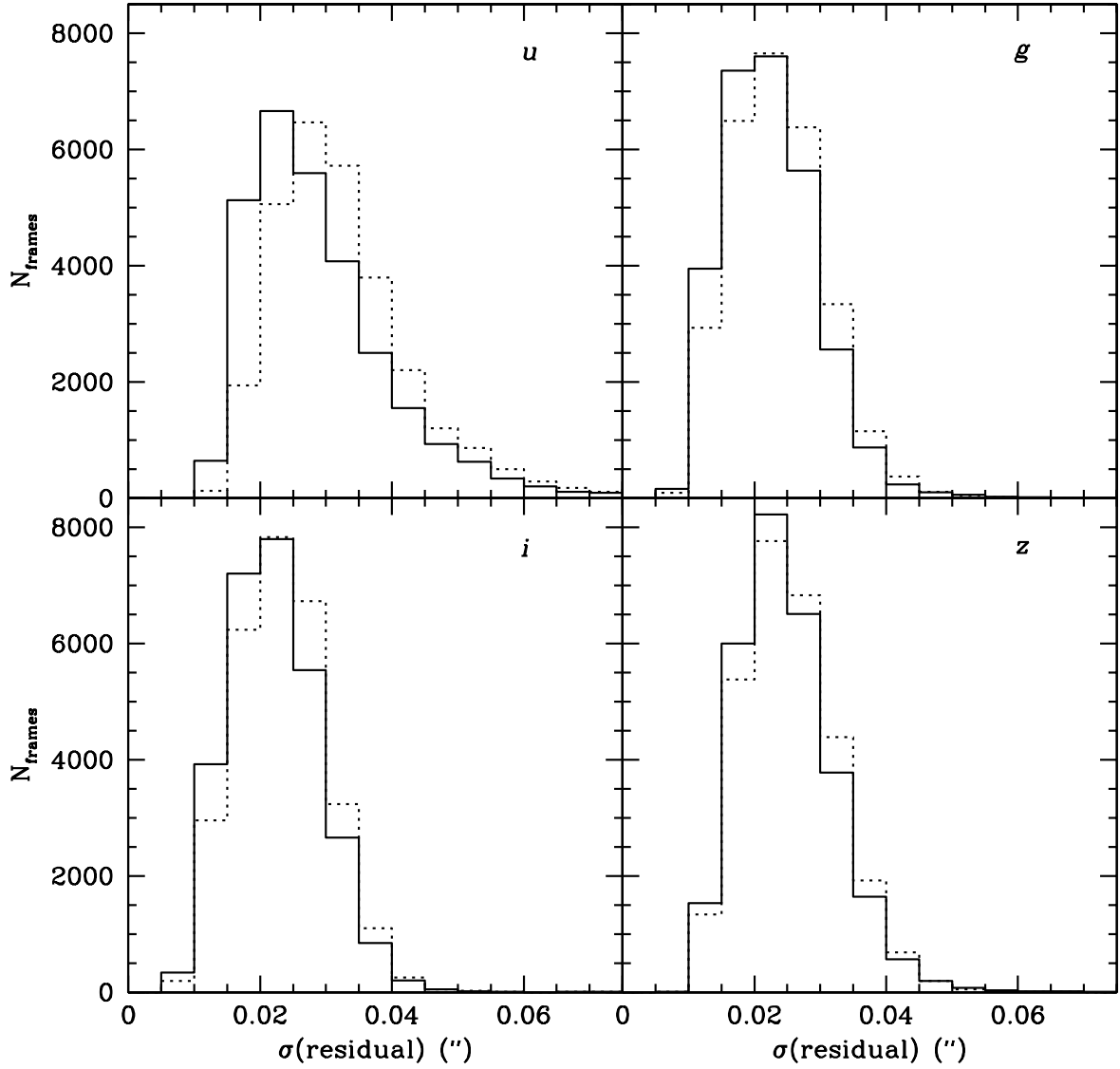}
\caption{Histograms of the rms residuals (SDSS position in the
specified filter minus SDSS position in the $r$ filter)
for each frame in all DR1 scans.  The rms residuals for each frame are
calculated using a minimum of the nearest 100 matched pairs.
The solid and dotted histograms are for the $\mu$ and $\nu$ rms residuals,
respectively.}
\label{fig-internalRelRms}
\end{figure}

\begin{figure}
\plotone{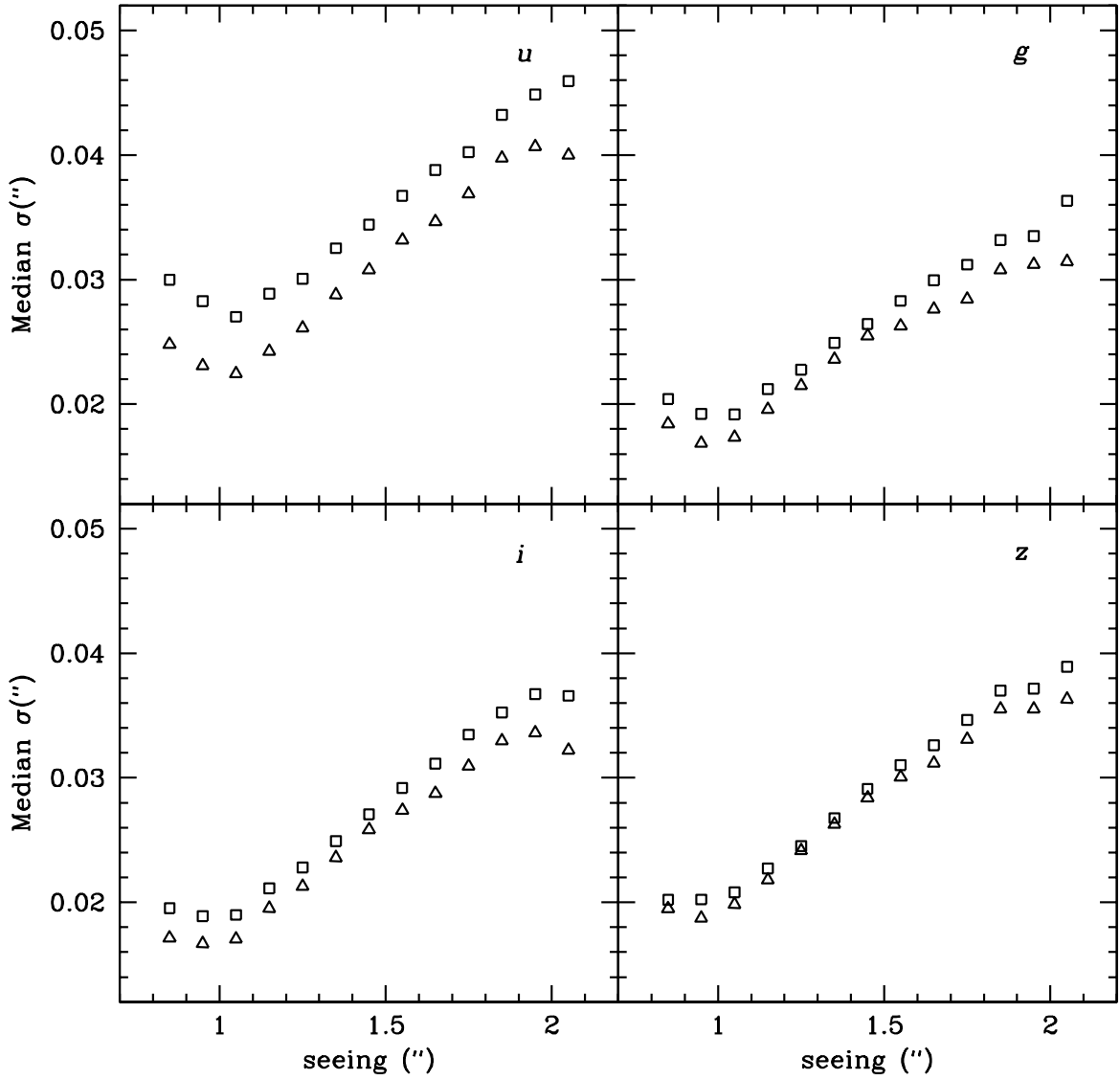}
\caption{Median rms residual per frame binned by seeing for the $u$, $g$, $i$,
and $z$ CCDs. Triangles are for the $\mu$ residuals, squares the $\nu$
residuals.}
\label{fig-relSeeing}
\end{figure}

The primary internal measure of the accuracy of the relative astrometry
between filters is the
distribution of residuals (measured positions in the
target filter -- $u$, $g$, $i$, or $z$ -- minus the measured positions in
the $r$ filter).
Figure~\ref{fig-internalRelDiff} shows histograms of these residuals
for all matched pairs (the relative
astrometry between filters is not sensitive to the choice of primary
reference catalog), separately for each filter.
The relative astrometry is best in the $g$ and $i$ filters, with rms residuals
of about 24 mas per coordinate.  The $z$ filter is
only slightly worse, with rms residuals of about 27 mas per coordinate.
The centroids are poorer in the $u$ filter due to the lower
flux for the stars used to derive the transformations, leading to larger rms
residuals of about 33 mas per coordinate.
Again, the variation in astrometric accuracy can be examined by calculating
rms values for each frame, for each CCD pair.
To assure good statistics, a smoothing window is used, using enough adjacent
frames to give a minimum of 100 matched pairs per frame.
The rms residuals in $\mu$ and $\nu$ for each frame are recorded with
the calibration equations for that frame, and serve as the best estimate
of the systematic error for that frame.
Figure~\ref{fig-internalRelRms} shows histograms of these rms residuals.
The distributions are fairly narrow in $g$, $i$, and $z$, with most frames
having rms residuals of less than 35 mas.  The distribution in $u$ shows a tail
out to 60 mas.  This is due both to the increased centroiding errors in
the stars used to calibrate the $u$ frames, as well as to an occasional lack of
enough bright stars to adequately calibrate the frame.
The rms residuals are sensitive to seeing, as shown in
Figure~\ref{fig-relSeeing}, which plots the median rms residual per frame
binned by seeing.

\begin{figure}
\plotone{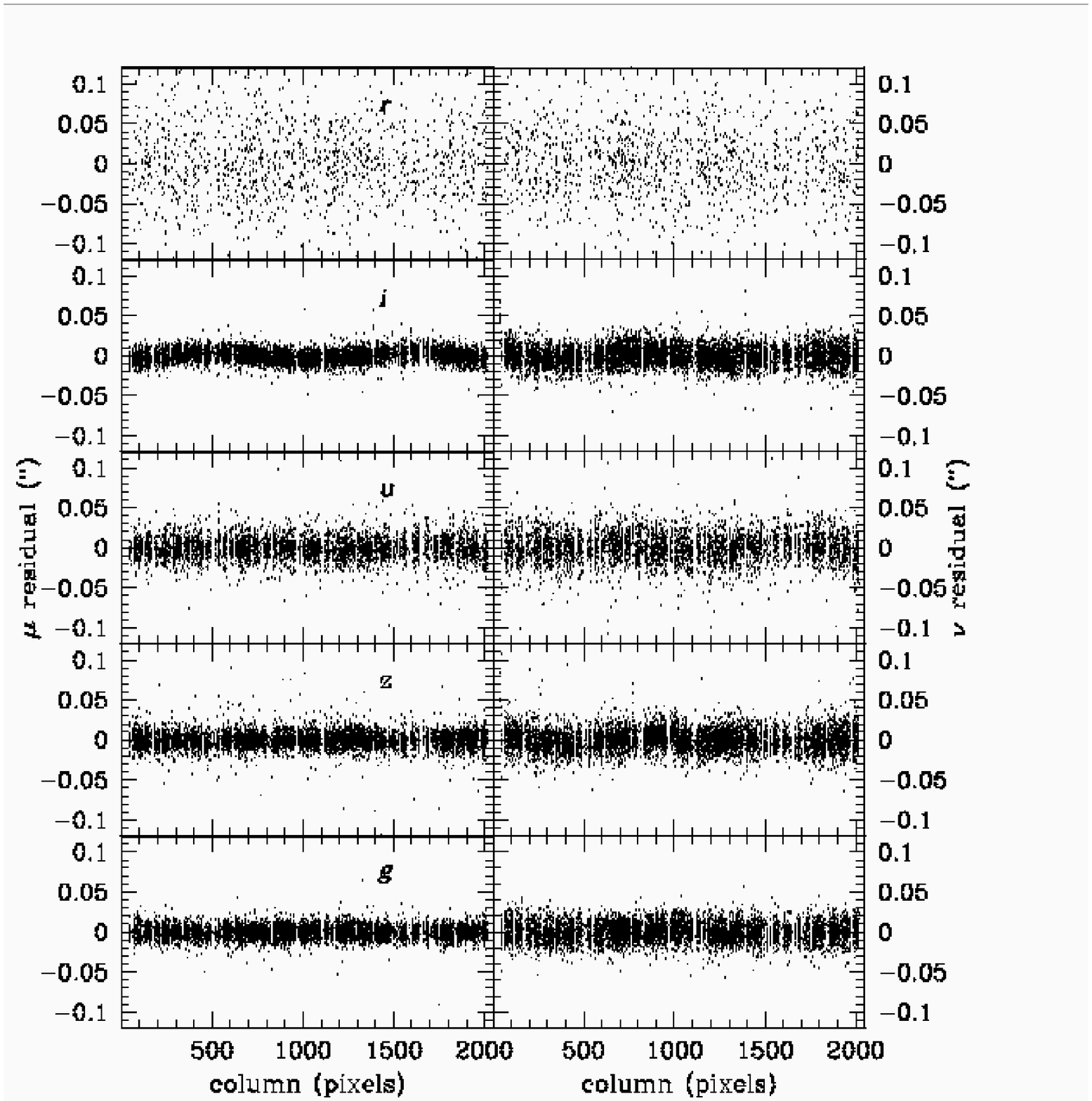}
\caption{$\mu$ and $\nu$ residuals plotted against CCD column for all
five photometric CCDs in column 3 of the camera for a scan taken under
exceptional seeing.  The scatter for the $r$ CCD is larger than the other CCDs
because $r$ has been matched against UCAC while the other CCDs were matched
against the $r$ detections.}
\label{fig-column}
\end{figure}

\begin{figure}
\plotone{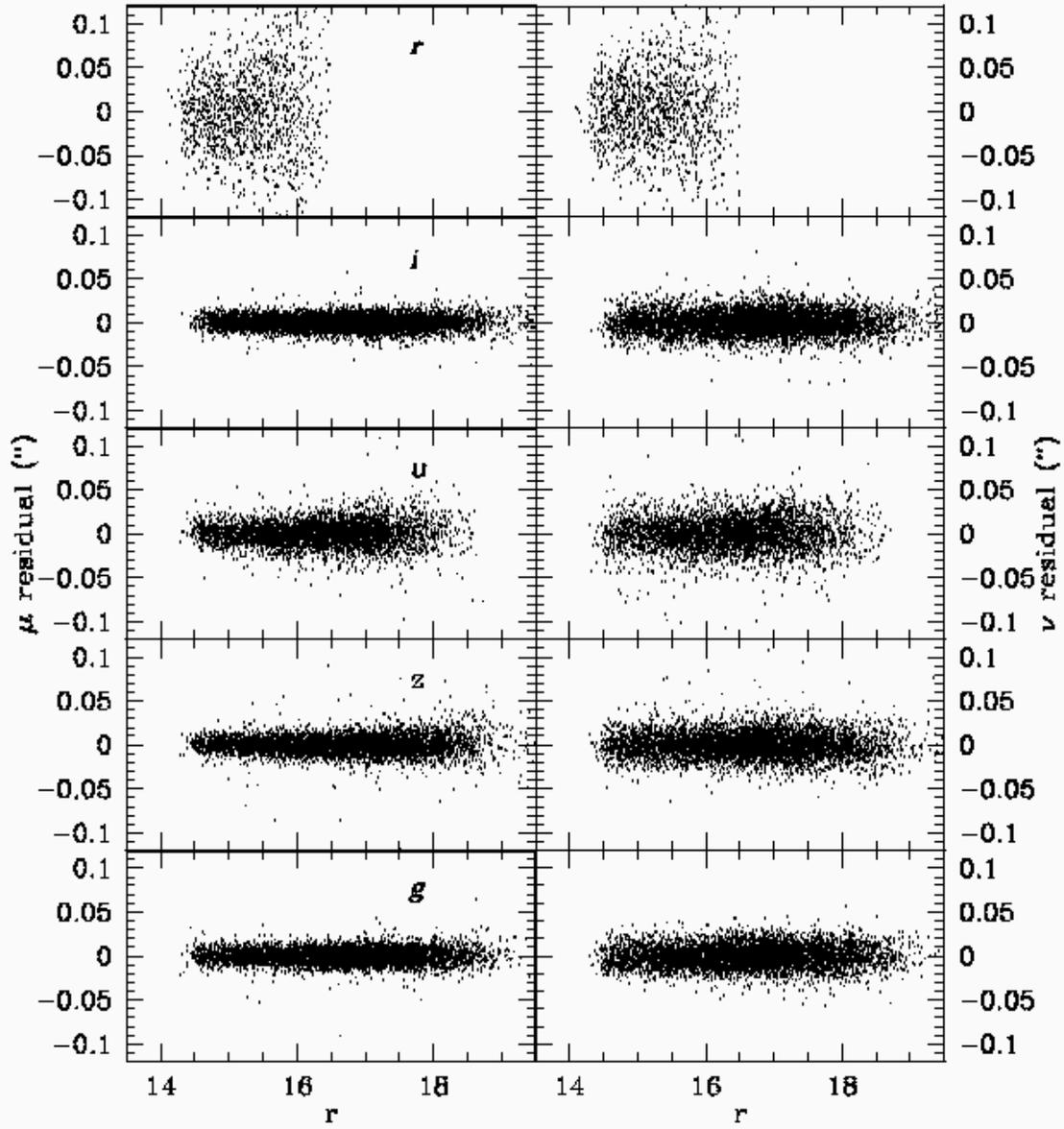}
\caption{Same as Figure~\ref{fig-column}, but plotted against magnitude.}
\label{fig-mag}
\end{figure}

\begin{figure}
\plotone{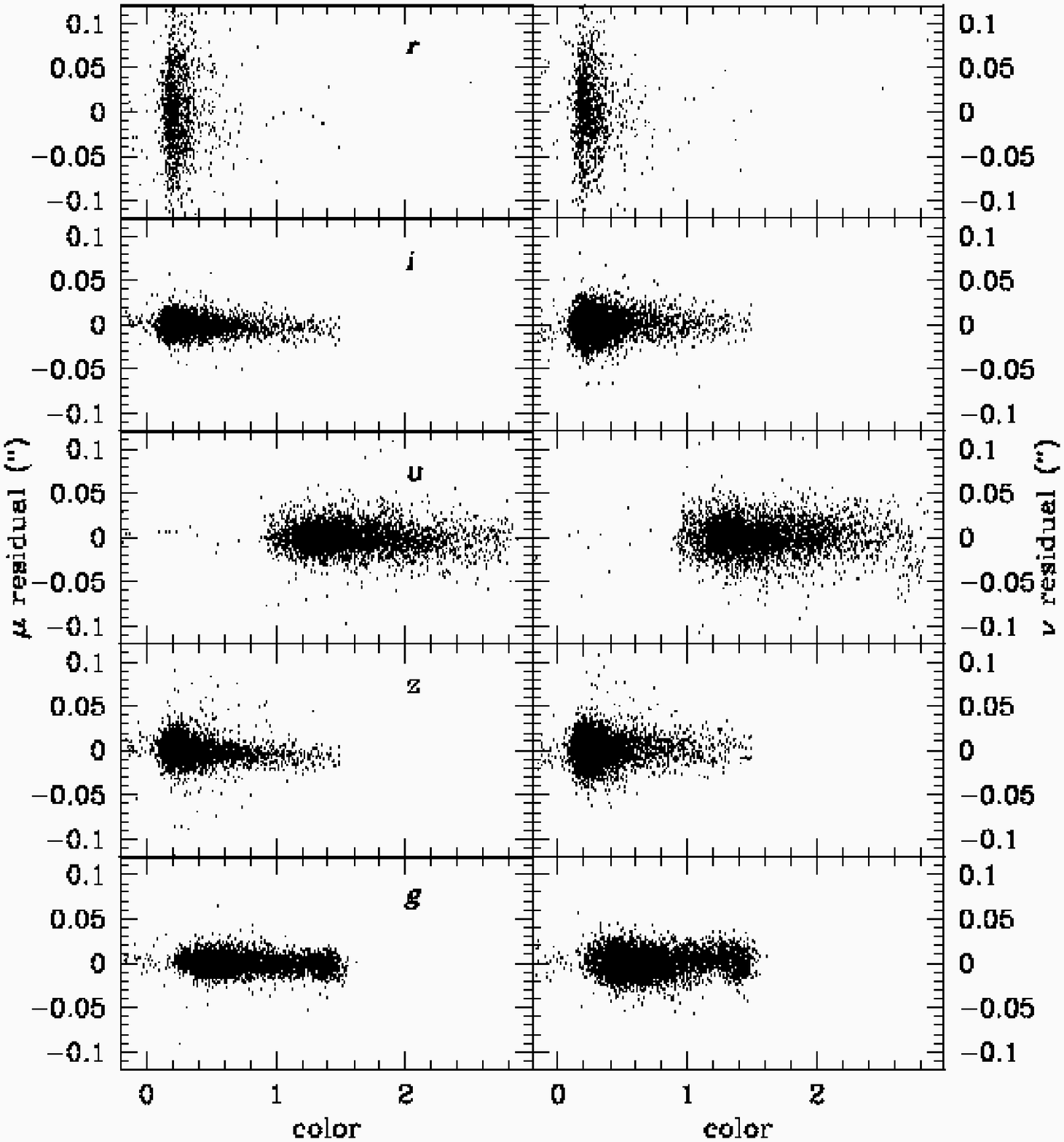}
\caption{Same as Figure~\ref{fig-column}, but plotted against color
($r - i$ for $r$, $i$, and $z$, $u - g$ for $u$, and $g - r$ for $g$.}
\label{fig-color}
\end{figure}

Figures~\ref{fig-column}, \ref{fig-mag}, and \ref{fig-color} plot the $\mu$
and $\nu$ residuals against CCD column, magnitude, and color, respectively, for
all five photometric CCDs in column 3 of the camera for a scan taken under
exceptional seeing to minimize the random errors and thus make any
remaining systematic errors more obvious.  The scatter for the $r$ CCD is larger
than the other CCDs because $r$ has been matched against UCAC while the other
CCDs were matched against the $r$ detections.
Systematics with column (uncorrected by the cubic polynomials used to remove
the optical distortions) are almost always less than 10 mas (though for
some $u$ CCDs they can peak at 15 mas), and typically are 5 mas or less.
Systematics with magnitude are always less than 13 mas over the magnitude
range $14 < r < 22$, and are typically less than half that.
Systematics with color are evident only for the $u$
and $g$ CCDs, and are typically less than 10 mas over the full range
in color.  All of these uncorrected systematics contribute to the rms
residuals quoted above.  

\subsection{Centroiding Errors}

\begin{figure}
\plotone{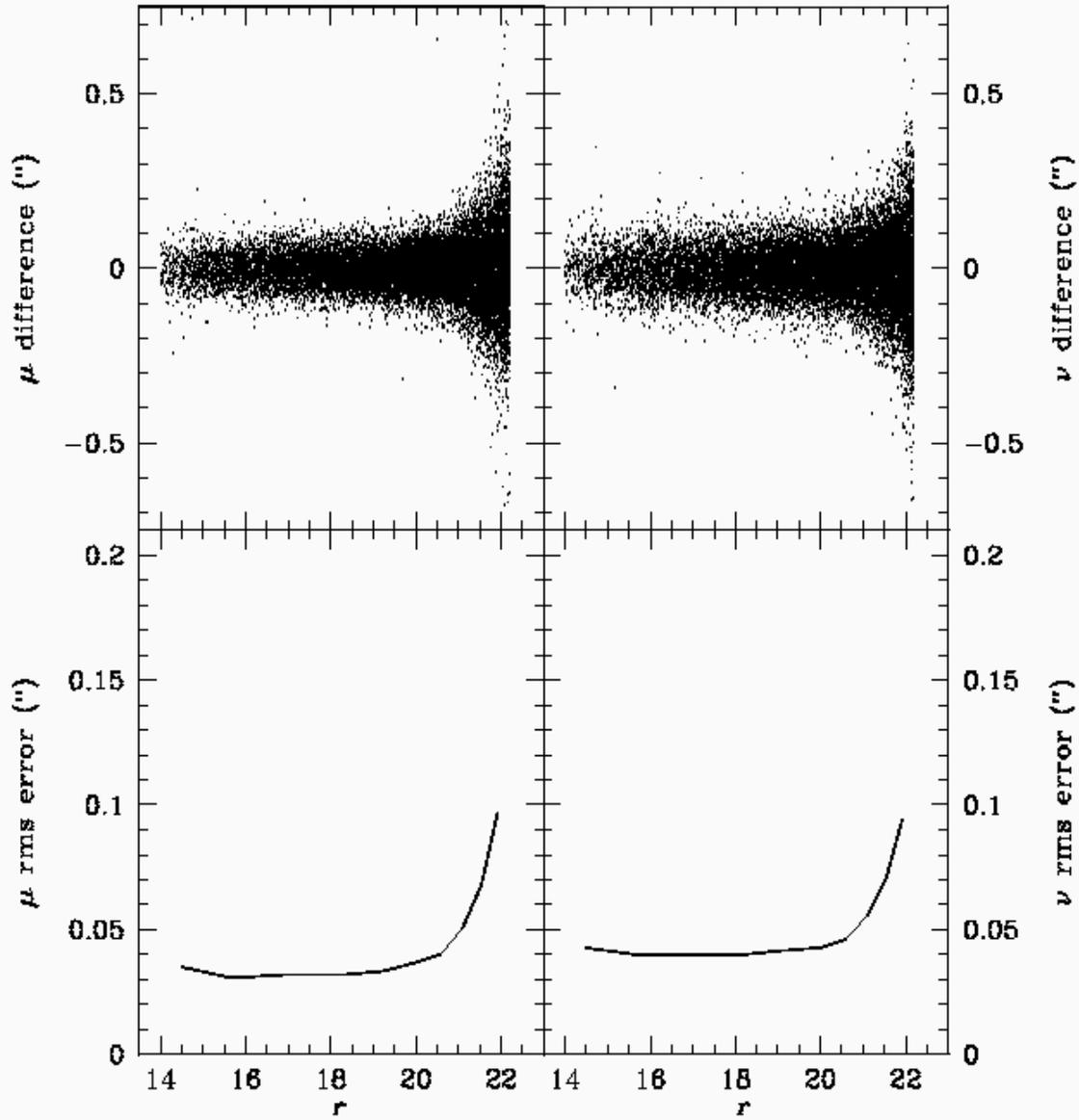}
\caption{Position differences for stars plotted against $r$ for two overlapping
scans (upper panels), and the rms position errors versus $r$ for stars for a
typical scan (taken under $1.4\arcsec$ seeing), calculated from the position
differences (lower panels).}
\label{fig-absFaintStars}
\end{figure}

The results presented above apply to stars brighter than $r \sim 20$,
where systematic errors dominate centroiding errors, and
thus characterize the errors in the calibrations.
Centroiding errors become significant for stars fainter than
$r \sim 20$, and for all galaxies.
Figure~\ref{fig-absFaintStars} shows, in the upper two panels,
the position differences against $r$ for matched stars between two overlapping
scans reduced against UCAC.  Each scan was taken under fairly constant
seeing of $1.4\arcsec$ (the median seeing for all DR1 data is about
$1.4\arcsec$).  The errors clearly increase for stars fainter
than $r \sim 20$.  The position errors as a function of magnitude for a
run taken during seeing of $1.4\arcsec$ can be estimated by binning
the position differences in magnitude and calculating the rms differences in
each magnitude bin, and dividing the rms by $\sqrt{2}$.  This has been done in
the lower panels of Figure~\ref{fig-absFaintStars}.  The rms errors increase
from about 40 mas at $r \sim 20$ to about 100 mas at the survey limit of
$r \sim 22$.  The errors are seeing dependent, with rms errors at
$r \sim 22$ ranging from about 70 mas to 140 mas in seeing of approximately
$1.1\arcsec$ and $1.7\arcsec$, respectively (a range of seeing which
characterizes most of the DR1 data).

\begin{figure}
\plotone{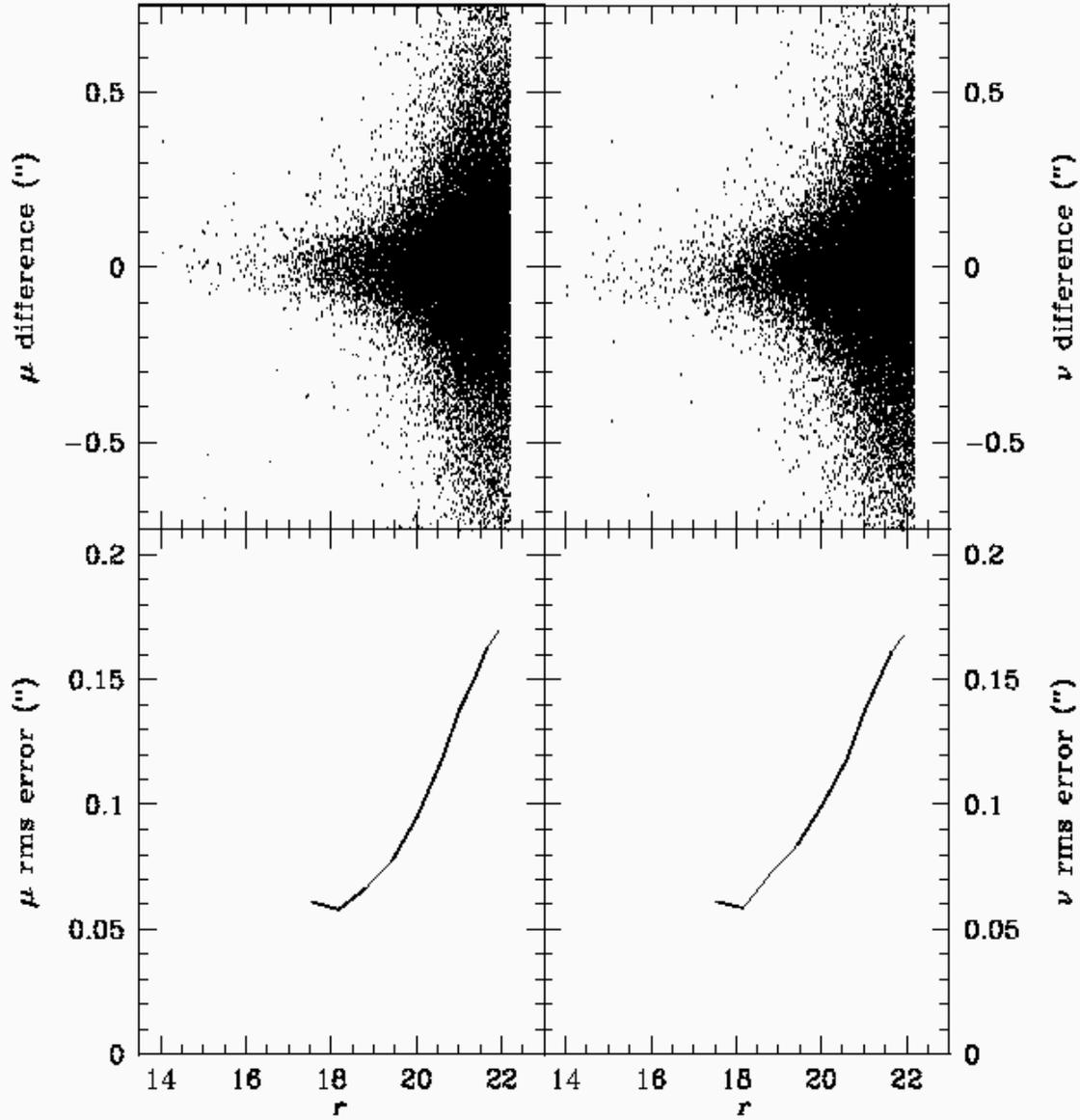}
\caption{Same as Figure~\ref{fig-absFaintStars}, but for galaxies.}
\label{fig-absFaintGalaxies}
\end{figure}

\begin{figure}
\plotone{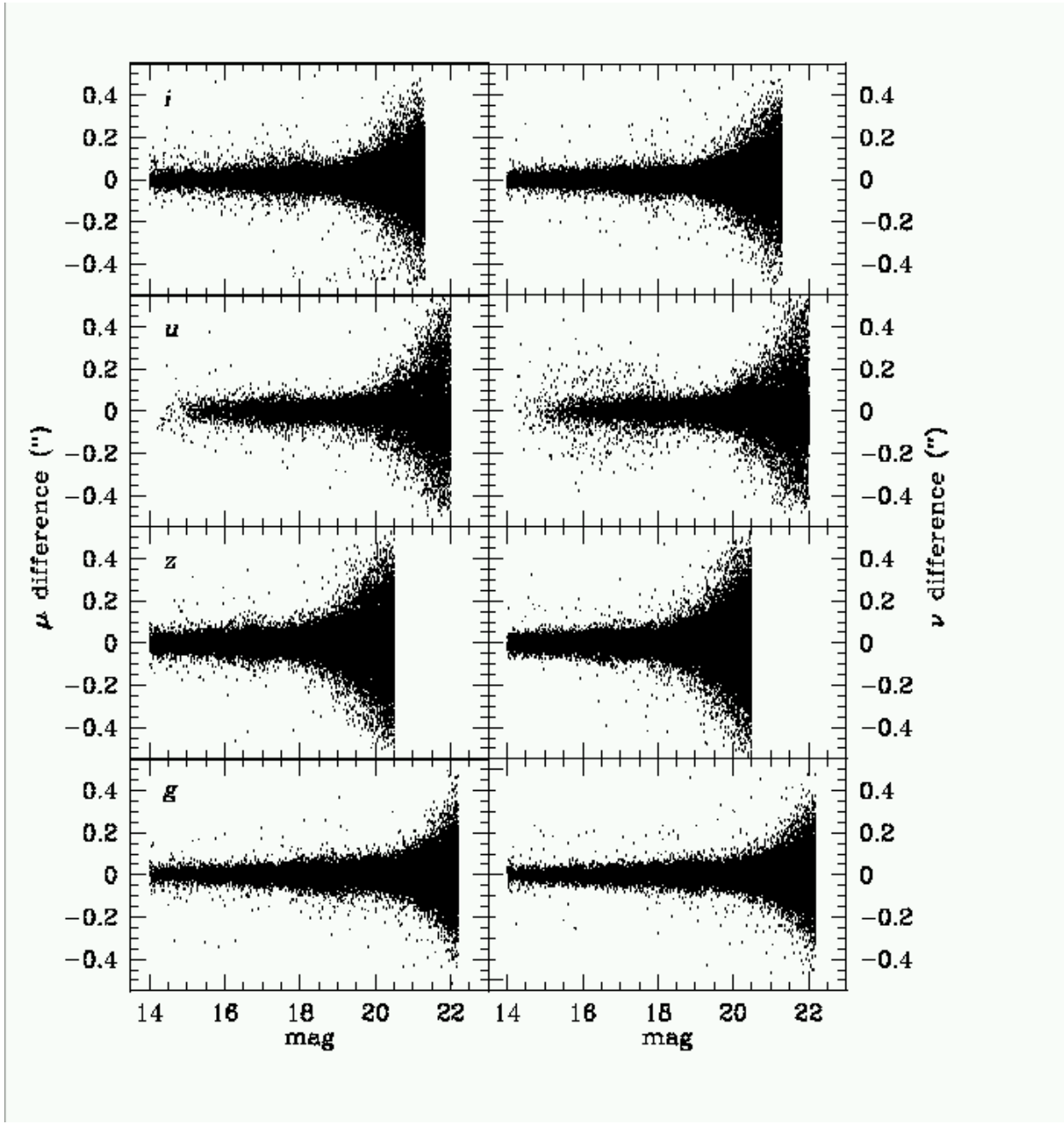}
\caption{Position differences between the
$i$, $u$, $z$, and $g$ filters, and the $r$ filter,
plotted against magnitude (\iuzg) for a typical scan.}
\label{fig-relFaintScatter}
\end{figure}

\begin{figure}
\plotone{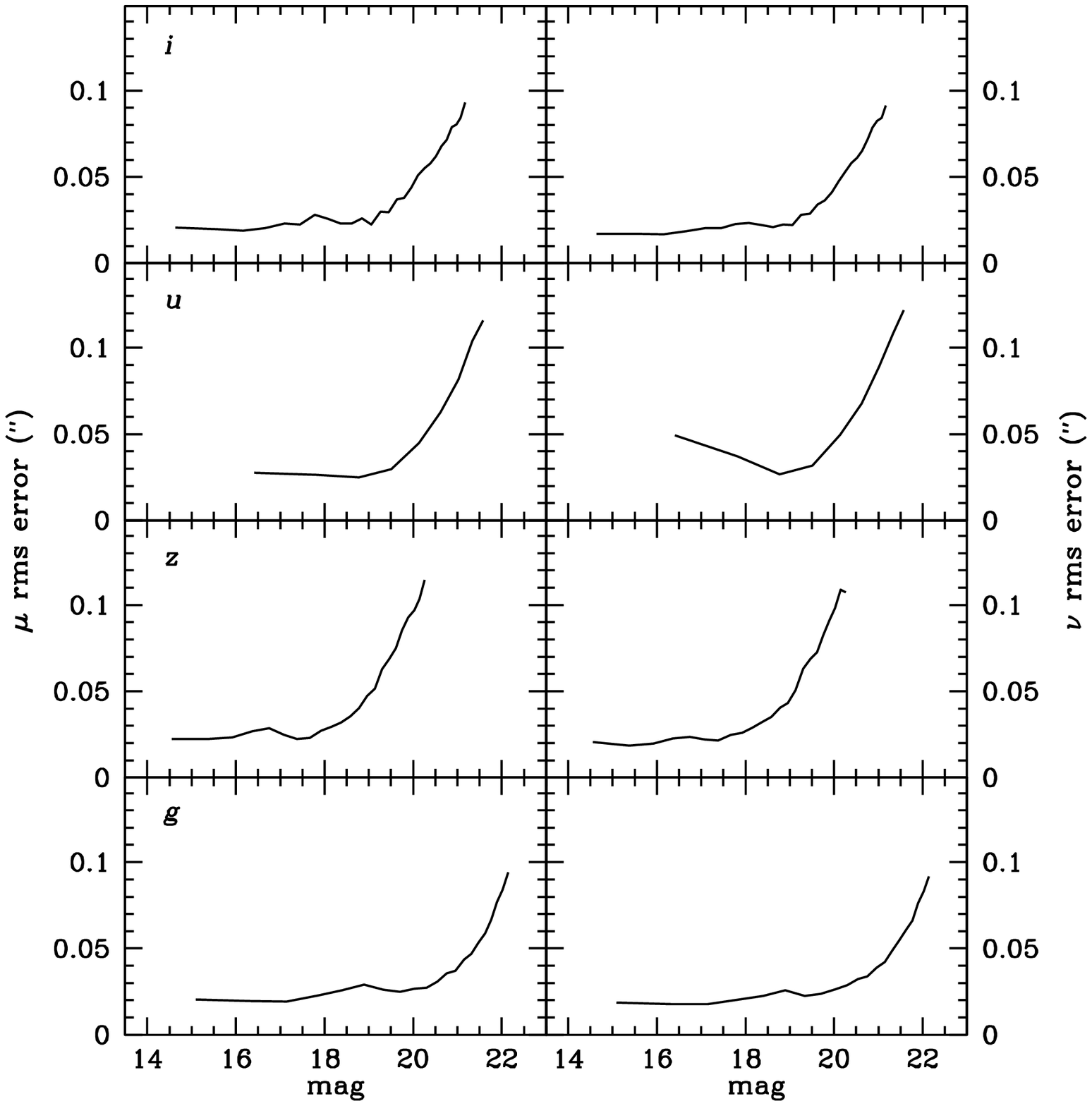}
\caption{Rms errors for the relative astrometry between the
$i$, $u$, $z$, and $g$ filters, and the $r$ filter,
plotted against magnitude (\iuzg) for a typical scan.}
\label{fig-relFaintError}
\end{figure}

Figure~\ref{fig-absFaintGalaxies} is similar to Figure~\ref{fig-absFaintStars},
but plots the position differences and estimated errors for galaxies rather
than stars (for the same scans).  Centroiding errors dominate the calibration
errors even at bright magnitudes, with the rms errors ranging from about
60 mas at $r \sim 17$ to 170 mas at $r \sim 22$.  These results are
independent of seeing.

The accuracy of the relative astrometry between filters similarly decreases
with increasing magnitude.  Figure~\ref{fig-relFaintScatter} plots the position
differences between the $i$, $u$, $z$, and $g$ filters and the $r$ filter
against magnitude (\iuzg) for a single scan observed in $1.65\arcsec$ seeing.
The rms errors as a function of magnitude can be calculated for this scan
by binning the differences in magnitude and calculating the rms differences in
each magnitude bin.  These errors are shown in Figure~\ref{fig-relFaintError}.
The rms errors increase to around 100 -- 130 mas by the survey limits.

\subsection{Anomalous Refraction}
\label{sec-anomalous}

The primary contributions to systematic astrometric residuals appear to come
from one or both of two causes: atmospheric fluctuations and/or telescope
jitter.  These observed systematics display considerable coherence
across the focal plane; that is, both  the direction and amplitude of the
residuals from CCD to CCD show a high degree of correlation. These effects can
be seen in Figures~\ref{fig-run-to-run-time-ucac} and
\ref{fig-run-to-run-time-tycho2}.
The focal plane subtends $2.3\arcdeg$ across the sky and the physical dimension
is on the order of $50 $ cm (i.e., several times typical values for $r_0$).

Thus, one is led to believe that either the focal plane itself is moving
(due, e.g., to telescope jitter) or the atmosphere is coherently moving
the telescope beam.  These two effects are difficult to disentangle, and
it is with some caution that we assign these
effects to the atmosphere for the following reasons.

Among the factors one might imagine could cause the telescope to 
be the culprit are mechanical resonances, servo motor errors and/or bearing
unevenness which result in tracking errors, wind-induced motions, thermal
oscillations, and oscillations in the mirror support system.
Time scales for the systematic effects vary from night to night (and,
occasionally, from hour to hour), but are typically on the order of a few
minutes to a few tens of minutes.  The telescope mechanical resonances are
on the order of a few hertz or higher and would average out over an
integration time.  These same effects have been seen on engineering scans
during which the telescope was clamped on the meridian and equator with all
telescope drive motors turned off.  During such observations, no significant
motion of either the telescope mechanicals or optics, or of the imaging camera
and/or its CCDs, is thought to have contributed to the effects during
these scans.  Hence we are persuaded that neither tracking
errors nor mirror motions cause these effects.  Thermal time scales are
much longer (an hour or more) and thus also seem to be an unlikely cause.
Wind induced motions remain a possibility, though we see such effects on
all of our scans, no matter the wind speed or direction, and the telescope
is enshrouded in a mechanically isolated wind baffle to prevent just such
a problem.

Further, this kind of behavior has been seen before in drift scanning CCD
observations.  \citet {dunham} observed several strips multiple times with
a scanning CCD. A comparison of the residuals ``revealed the presence of
a pervasive low-frequency motion of the sky coordinate system relative to
the CCD." They also found that the variations along right ascension and
declination were significantly
different.  \citet {stoneanom} reported similar results with their drift-scan
mode CCD observations made on an extremely stable transit telescope.  Both
of these sets of observations were obtained with single CCDs (a couple of
cm in size at most) with fields of view of about $20'$.
\citet {stoneanom} attribute the cause of these effects to
{\em anomalous refraction}, i.e., refraction that varies from the smooth
analytical models of refraction which are functions only of zenith distance.

The SDSS observations are, we believe, the first astronomical observations to
demonstrate that these effects are due to phenomena whose scale is on the order
of two degrees or more on the sky.  \citet {webb} and \citet {naito} discuss
atmospheric boundary layers with heights of typically a few hundred meters
to two km above the earth's surface and horizontal wavelengths ($\lambda$)
of $1$ to $10$ or more km which may give rise to anomalous refraction.
If this is true, then we would expect to see characteristic time scales of
$\lambda/v$ (where $v$ is the wind speed).  For typical values ($1 \le \lambda
\le 10$ km, and $10 \le v \le 40$ km hour$^{-1}$),
we expect to see time signatures
of a few to several tens of minutes, in agreement with the SDSS
observations.

\section{Early Data Release}
\label{sec-edr}

This paper describes the astrometric calibration procedure based on version
v3\_6 of \astrom\ and version v5\_3 of \frames.
On June 5, 2001, the SDSS made its first public data release of primarily
commissioning data, referred to as the Early Data Release
\citep[EDR,][]{edr}.  The EDR data
were processed with earlier versions of both \astrom\ and \frames,
and so the astrometric calibration procedure differed to some extent from the
procedure described in this paper.  The primary difference for astrometry
concerns the centroids, which were not corrected for asymmetries in the PSFs.
More significantly,
\frames\ used a different smoothing length, as well as different
magnitude-dependent binning factors, than did \ssc\ (earlier versions of\
\astrom\ used
uncorrected centroids from \ssc, rather than the currently corrected centroids
from \psp) before calculating the
centroids.  This led to systematic offsets between the \ssc\ and \frames\
centroids of up to 20 mas, as well as systematic offsets between \frames\
centroids for bright and faint objects of order 20 mas (the break occurs
around $r \sim 20$).  Since the calibrations were based on \ssc\ centroids,
but the final positions are based on \frames\ centroids,
this contributed an additional 20 mas of systematic error in the EDR data for
bright ($r \lesssim 20$) objects, and perhaps twice that for faint
objects.  Roughly 90\% of the data in the EDR have been
recalibrated using the new procedure, and these new calibrated positions will
be included in DR1.

\section{Summary}
\label{sec-conclusion}

The astrometric calibration of the SDSS uses two reduction strategies.
In those areas of the sky with UCAC coverage, the $r$ CCDs are calibrated
directly against UCAC.  In those areas of the sky which lack UCAC coverage,
bright stars detected on the astrometric CCDs are transferred to the pixel
coordinate systems of the $r$ CCDs
using stars in common to both CCDs, and the bright stars are then used to
calibrate the $r$ CCDs against Tycho-2.
The accuracy of the calibrations using UCAC are of order 45 mas rms, with an
additional systematic error of up to 30 mas (due primarily
to systematic errors in UCAC).
The accuracy of calibrations using Tycho-2 are of order 75 mas with
an additional systematic error of order 20 mas
(due to CTE effects in the astrometric CCDs).  The rms errors are dominated
by Gaussian distributions of systematic errors which vary on time scales of
one to a few tens of minutes due to anomalous refraction,
and by random errors in the primary reference catalogs.
The accuracy of the relative astrometry of the
$u$, $g$, $i$, and $z$ filters versus the $r$ filter is of order 25 mas rms
for the $g$, $i$, and $z$ filters, and 35 mas rms
for the $u$ filter.  Systematic errors with magnitude, color, or CCD column
are typically less than 10 mas.

The astrometric performance being realized by SDSS far exceeds the
original design specifications and science requirements.  There are a
number of reasons for this:
\begin{enumerate}
\item  The imaging camera has proven to be rigid and stable and makes
no significant contribution to astrometric errors.  It was originally
anticipated that dewar-to-dewar excursions of a few microns (the scale
is $\approx  60\ \mu$ per arcsec) and an undetermined though small amount of
flexure might need to be tolerated.  The camera's performance has been
exceptional.
\item  The telescope tracking has proven to be extremely accurate and reliable.
Considerable engineering and design effort went into optimizing the SDSS
2.5m telescope performance, with obvious benefits for astrometry and
imaging quality.
\item  The availability of new, deeper, and better astrometric catalogs
(e.g., Tycho-2 and UCAC) has made a significant impact on improving the
astrometry.
\end{enumerate}

UCAC provides a valuable densification of the Hipparcos Reference Frame (HRF),
the best current optical realization of the International Celestial Reference
Frame (ICRF), extending it to $R \sim 16$.  Ultimately, the SDSS will provide,
for about a fourth of the sky, a further densification of the HRF, providing
astrometry relative to the HRF (via its calibration against UCAC) accurate
to 45 mas rms plus 30 mas systematic to $r \sim 20$, and approximately
100 mas rms at $r \sim 22$.

The accuracy of the astrometry relative to the HRF allows for precise matching
with other catalogs which have also been calibrated within the HRF.  Various
proper motion studies are currently underway based on matches to USNO-B
\citep{usnob}.  Matches between the SDSS and 2MASS
have been used to study the optical and infrared photometric properties of
stars \citep{finlator} and to search for $z > 5.8$ quasars \citep{qsos}.
Matches between the SDSS and the Faint Images of the Radio Sky at Twenty-cm
\citep[FIRST,][]{first} survey are being used to study the optical
and radio properties of extragalactic sources \citep{ivezic}.
The ease with which such studies can be made is directly attributable to
the impact of the watershed Hipparcos space astrometry mission
\citep{hipparcos} on global astrometry.
With the accuracy of the relative astrometry between filters, asteroids are
easily detected based on their relative motions between filters within
a single scan \citep{asteroids}, and the same technique offers promise for the
detection of Kuiper Belt Objects.

\acknowledgements
Funding for the creation and distribution of the SDSS Archive has been
provided by the Alfred P. Sloan Foundation, the Participating Institutions,
the National Aeronautics and Space Administration, the National Science
Foundation, the U.S. Department of Energy, the Japanese Monbukagakusho, and
the Max Planck Society. The SDSS Web site is http://www.sdss.org/. 

The SDSS is managed by the Astrophysical Research Consortium (ARC) for the
Participating Institutions. The Participating Institutions are The University
of Chicago, Fermilab, the Institute for Advanced Study, the Japan
Participation Group, The Johns Hopkins University, Los Alamos National
Laboratory, the Max-Planck-Institute for Astronomy (MPIA), the
Max-Planck-Institute for Astrophysics (MPA), New Mexico State University,
University of Pittsburgh, Princeton University, the United States Naval
Observatory, and the University of Washington.

Fermilab is operated by Universities Research Association Inc. under
Contract No. DE-AC02-76CH03000 with the United States Department of
Energy.
RHL and ZI would like to thank Princeton University for generously providing
research funds, and Jill Knapp for generously providing lunch and
sympathy.  We thank the entire UCAC team for making available to the
SDSS pre-release versions
of UCAC, Mario Nonino for work on an early version of the pipeline,
Walter Siegmund for useful discussions, and N. Zacharias, D. York, and
M. Strauss for critical readings of the manuscript.
\astrom\ makes extensive use of the SLALIB library \citep{slalib} from the
Starlink Project, which is run by CCLRC on behalf of PPARC.
This publication makes use of data products from the Two Micron All Sky
Survey, which is a joint project of the University of
Massachusetts and the Infrared Processing and Analysis Center/California
Institute of Technology, funded by the National
Aeronautics and Space Administration and the National Science Foundation. 
This research has made use of the NASA/IPAC Infrared Science Archive, which
is operated by the Jet Propulsion Laboratory, California Institute of
Technology, under contract with the National Aeronautics and Space
Administration.

\clearpage

\end{document}